\documentclass[twocolumn,showpacs,aps,amsmath,amssymb,superscriptaddress,prb]{revtex4-1}
\usepackage{amsfonts}
\usepackage{amsmath}
\usepackage{mathtools, amssymb}
\usepackage{bm}
\usepackage{graphicx}
\usepackage{overpic}
\usepackage{color}
\usepackage{multirow}
\usepackage{tikz-cd}
\usepackage{float}
\usepackage{epstopdf}
\usepackage{url}
\usepackage{array}
\usepackage{setspace}
\allowdisplaybreaks
\newcommand{\Sec}[1]{Sec.\,\ref{#1}}

\newcommand{\nl}{\nonumber \\}
\newcommand{\be}{\begin{equation}}
\newcommand{\ee}{\end{equation}}
\newcommand{\bea}{\begin{eqnarray}}
\newcommand{\eea}{\end{eqnarray}}
\newcommand{\Fig}[1]{Fig.~\ref{#1}}

\newcommand{\Eq}[1]{Eq.\,(\ref{#1})}
\newcommand{\Eqs}[1]{Eqs.\,(\ref{#1})}
\newcommand{\Sch}{Schr\"{o}dinger\ }
\newcommand{\la}{\langle}
\newcommand{\ra}{\rangle}
\newcommand{\w}{\omega}
\newcommand{\Ht}{H_{_{\rm T}}}
\newcommand{\Hs}{H_{_{\rm S}}}
\newcommand{\Hb}{H_{_{\rm B}}}
\newcommand{\Hsb}{H_{_{\rm SB}}}
\newcommand{\rhos}{\rho_{_{\rm S}}}
\newcommand{\rhob}{\rho_{_{\rm B}}}
\newcommand{\rhot}{\rho_{_{\rm T}}}
\newcommand{\trhos}{\tilde{\rho}_{_{\rm S}}}
\newcommand{\brhos}{\bar{\rho}_{_{\rm S}}}
\newcommand{\drhob}{\dot{\rho}_{_{\rm B}}}
\newcommand{\dtrhos}{\dot{\tilde{\rho}}_{_{\rm S}}}
\newcommand{\iChEM}{{\it i}{\rm ChEM}}%
\begin{document}

\title{Stochastic Equation of Motion Approach to Fermionic Dissipative Dynamics. II. Numerical Implementation}

\author{Arif Ullah}
\affiliation{Hefei National Laboratory for Physical Sciences at the Microscale \&
	Synergetic Innovation Center of Quantum Information and Quantum Physics,
	University of Science and Technology of China, Hefei, Anhui 230026, China}

\author{Lu Han}
\affiliation{Hefei National Laboratory for Physical Sciences at the Microscale \&
	Synergetic Innovation Center of Quantum Information and Quantum Physics,
	University of Science and Technology of China, Hefei, Anhui 230026, China}

\author{Yun-An Yan}
\affiliation{School of Physics and Optoelectronic Engineering, Ludong University, Shandong 264025, China}

\author{Xiao Zheng} \email{xz58@ustc.edu.cn}
\affiliation{Hefei National Laboratory for Physical Sciences at the Microscale \&
	Synergetic Innovation Center of Quantum Information and Quantum Physics,
	University of Science and Technology of China, Hefei, Anhui 230026, China}

\author{YiJing Yan}
\affiliation{Hefei National Laboratory for Physical Sciences at the
	Microscale \& \iChEM, University of Science and Technology of China, Hefei, Anhui
	230026, China}

\author{Vladimir Chernyak}
\affiliation{Hefei National Laboratory for Physical Sciences at the Microscale \&
	Synergetic Innovation Center of Quantum Information and Quantum Physics,
	University of Science and Technology of China, Hefei, Anhui 230026, China}
\affiliation{Department of Chemistry, Wayne State University, 5101 Cass Avenue, Detroit, MI 48202}

\date{Submitted on December~10, 2019}

\begin{abstract}

This paper provides a detailed account of the numerical implementation of the stochastic equation of motion (SEOM) method
for the dissipative dynamics of fermionic open quantum systems.
To enable direct stochastic calculations, a minimal auxiliary space (MAS) mapping scheme is adopted,
with which the time-dependent Grassmann fields are represented by c-numbers noises and a set of pseudo-operators.
We elaborate on the construction of the system operators and pseudo-operators involved in the MAS-SEOM,
along with the analytic expression for the particle current.
The MAS-SEOM is applied to study the relaxation and voltage-driven dynamics of 
quantum impurity systems described by the single-level Anderson impurity model, 
and the numerical results are benchmarked against those of the 
highly accurate hierarchical equations of motion (HEOM) method.
The advantages and limitations of the present MAS-SEOM approach are discussed extensively.

\end{abstract}

\maketitle

\section{Introduction} \label{sec:intro}

Fermionic dissipative system refers to a quantum system
embedded in or coupled to a fermionic environment,
and exchanges energy, particles, and/or quantum phase with it.
A well-known example of fermionic dissipative system is the quantum impurity system (QIS),
which normally consists of one or more quantum impurities (core system)
and a number of electron reservoirs (surrounding environments).
The QIS is of fundamental interest and importance to many fields of physics, chemistry, and material sciences.
For instance, QIS such as quantum dots\cite{rasanen2004impurity,Zhe13086601,Ye14165116,Hou15104112,gong2018quantum}  
and molecular magnets\cite{Hei154024,heinrich2018control,wang2018precise,czap2019probing,coronado2019molecular,walkey2019chemically}
have found many applications, including quantum control,\cite{d2007introduction}
quantum information storage\cite{song2005quantum} and processing,\cite{y2016principles}
and quantum computation.\cite{Leu01789,b2006quantum}
For practical purposes, it is crucial to understand how the fermionic environments,
such as the electron reservoirs, influence the local quantum states in the QIS.

Enormous efforts have been devoted to achieving an accurate characterization of QIS.
A variety of theoretical methods have been developed.
These include the numerical renormalization group (NRG) method,\cite{wilson1975renormalization,bulla2008numerical}
the density matrix renormalization group (DMRG) method,\cite{white1992density}
the exact diagonalization (ED) method,\cite{caffarel1994exact}
the quantum Monte Carlo (QMC) method,\cite{hirsch1986monte,gull2011continuous,cohen2015taming,antipov2016voltage,ridley2019lead}
the multi-configuration time-dependent Hartree (MCTDH) method\cite{meyer1990multi}
and its extensions,\cite{wang2003multilayer,Wan09024114}
and the iterative summation of real-time path integral (ISPI) method.\cite{weiss2008iterative,muhlbacher2008real,segal2010numerically}
Despite their success, these methods still have their respective limitations
in accuracy, efficiency or applicability.\cite{ridley2019lead} 

Besides the above methods, another popular method is the hierarchical equations of motion (HEOM) method.\cite{tanimura1989time,tanimura1990nonperturbative,Yan04216,Xu05041103,jin2008exact,shi2009efficient,
Li12266403,Har13235426,Sch16201407,Erp18064106,shi2018efficient}  
The HEOM method is capable of capturing the combined effects of system-environment dissipation,
non-Markovian memory effect, and many-body correlation in a non-perturbative manner.\cite{zheng2009numerical}
Conventionally, a hierarchy of deterministic differential equations are constructed
by using a set of memory basis functions (such as exponential functions) to unravel
the reservoir correlation functions,
%
The size of the HEOM is thus determined by two parameters, $M$ and $L$,
where $M$ is the number of memory basis functions used,
and $L$ is the depth of the hierarchy which depends critically
on the strength of system-environment interaction and many-body correlation.
In the case of low temperature and strong dissipative interaction,
an accurate characterization of the static and dynamic properties of a QIS
requires a large $M$ and $L$, which inevitably makes the numerical
calculations using the HEOM method rather expensive.\cite{han2018exact}
Such a drawback has restrained the use of the HEOM method
in the ultra-low temperature regime.
%


An alternative approach to theoretically address the QIS is the
stochastic quantum dissipation theory.
The stochastic approach is potentially promising particularly for the ultra-low temperature regime,
in which the effects of many-body correlation are prominent.
In our preceding paper\cite{han2019fermionic} (referred to as paper~I),
we have established the stochastic equation of motion (SEOM) formalism
for describing the dissipative dynamics of fermionic open systems.
In this formalism, the dynamics of the system and the fermionic environment
are decoupled by introducing the stochastic auxiliary Grassmann fields (AGFs).\cite{han2019stochastic}
This results in a formally exact SEOM for the stochastic system reduced density matrix.
However, such a rigorous SEOM is numerically unfeasible because of
the difficulty in realizing the anti-commutative AGFs.\cite{suess2015hierarchical,hsieh2018unified}
To enable direct stochastic calculation, we have further proposed a
minimum auxiliary space (MAS) mapping scheme, with which the AGFs
are represented by stochastic c-number noises and a set of pseudo-levels.
This leads to a numerically feasible MAS-SEOM approach\cite{han2019stochastic,han2019fermionic}
that could be used straightforwardly to simulate the dissipative dynamics of a QIS.

In Paper~I, we have proved the formal equivalence between the fermionic
SEOM and HEOM formalisms.
After the MAS mapping, the MAS-SEOM is found to be equivalent to
a simplified version of HEOM (sim-HEOM).\cite{han2018exact}
Regarding the practical applications, the SEOM method does not require
an explicit unraveling of the reservoir correlation functions,
and thus its memory cost is much less than that of the HEOM.
Moreover, because of the highly connected structure of the hierarchy,
the parallel implementation of the HEOM method is nontrivial.
Nevertheless, in recent years, there have been many works on the
parallelization of the HEOM method, such as the GPU-HEOM.\cite{kreisbeck2011high,strumpfer2012open,tsuchimoto2015spins}
In contrast, the SEOM can be solved by generating a number of
mutually independent quantum trajectories, and thus it is easy
to implement parallel computational techniques by employing
the trajectory-based algorithms.

In a stochastic formulation, the influence of environment on the system
can be captured by introducing stochastic auxiliary fields.
In the case of a boson bath, the stochastic fields are easily realized by c-number noises.
Consequently, the bosonic SEOM has been established and adopted by many authors.
For instance, Stockburger \emph{et~al.} have implemented the bosonic SEOM
by using Gaussian color noises.\cite{stockburger1998dynamical,stockburger2001non,stockburger2002exact,koch2008non}
Shao and coworkers have constructed a bosonic SEOM with memory-convoluted noises,\cite{shao2004decoupling,zhou2005stochastic}
which can be generated by using the fast Fourier transform.\cite{shao2010rigorous}
They have further reduced the number of noises by introducing correlated color noises.\cite{yan2016stochastic}
These color noises can be generated by the circulant embedding method\cite{chan1999simulation,percival2005exact}
or the spectral method.\cite{ding2011efficient,yizhao2013nonmarkovian}
Yan and Zhou have further proposed a Hermitian SEOM to improve the numerical convergence.\cite{yan2015hermitian}
There have also been attempts to combined the merits of the stochastic and hierarchical approaches.
For instance, the hybrid stochastic and hierarchical equations of motion (sHEOM) methods
have been proposed by several authors.\cite{zhou2005stochastic,moix2013hybrid,zhu2013new}
The success of these approaches is due to the fact that it is easy to generate
stochastic c-number noises.

In contrast to the bosonic environment, the auxiliary fields
for the fermion reservoirs are Grassmann numbers (g-number)
which anti-commute with each other.
It would thus require $N$ mutually anti-commutative matrices of
the size $2^N \times 2^N$ to represent $N$ g-numbers.
This immediately becomes unfeasible as $N$ increases.\cite{dalton2014phase}  
Such a difficulty has prohibited the numerical implementation of
the fermionic SEOM.\cite{applebaum1984fermion,rogers1987fermionic, hedegaard1987quantum,
zhao2012fermionic,chen2013non,shi2013non,chen2014exact,suess2015hierarchical,hsieh2018unified}
In the Paper~I, we have proposed a MAS mapping scheme as follows,
\be \label{map_eq1}
\eta_{jt} \mapsto v_{jt} X^-_j , \quad \bar{\eta}_{jt} \mapsto v_{jt} X^+_j , \quad (j=1,2,3,\ldots)
\ee
Here, $\{\eta_{jt}, \bar{\eta}_{jt}\}$ are time-dependent g-numbers,
$\{v_{jt}\}$ are Gaussian white noises,
and $\{X^\pm_j\}$ are the pseudo-operators defined in the auxiliary space  $S_j$ of a pseudo-level.
The details about the MAS mapping as well as the derivation of the
resulting MAS-SEOM have been presented in the paper~I.\cite{han2019stochastic}

In this paper (paper~II), we give a detailed account on the numerical aspects of the MAS-SEOM approach.
We will apply the MAS-SEOM approach to study the non-equilibrium dissipative dynamics of
a single-level Anderson impurity model (AIM). In particular,
we will explore the time-dependent electron transport properties of the AIM,
e.g., the time evolution of electron occupation number and the electric current flow
into the electron reservoirs,
In addition, we will examine the accuracy and efficiency of the MAS-SEOM,
as well as its convergence with respect to various parameters.

The remainder of this paper is arranged as follows.
\Sec{sec:method} is devoted to an illustration of the numerical implementation of the MAS-SEOM.
In \Sec{sec:current}, we worked out the formula for calculating the electric current flow
into the coupled electron reservoirs.
The asymptotic behavior of the stochastic noises involved in the MAS-SEOM are discussed in \Sec{sec:stability}.
The numerical results are presented and elaborated in \Sec{sec:results},
followed by concluding remarks and perspectives given in \Sec{sec:conclusion}.

\section{Numerical Implementation of the MAS-SEOM method}  \label{sec:method}

\subsection{MAS-SEOM for a single-level AIM} \label{subsec:mas-seom}

The total Hamiltonian of a single-level AIM consists of three parts:
\be \label{HT-1}
  H_{_{\rm T}} =  \Hs + \Hb + \Hsb.
\ee
The impurity (system) is described by the following Hamiltonian
(hereafter we adopt the atomic units $e = \hbar \equiv 1$ and $k_{\rm B} \equiv 1$):
\begin{equation}
\Hs= \epsilon_{\uparrow} \hat{n}_\uparrow + \epsilon_{\downarrow}\hat{n}_\downarrow + U\hat{n}_\uparrow \hat{n}_\downarrow .
\end{equation}
Here, $\hat{n}_s=\hat{c}^\dagger_s\, \hat{c}_s$ ($s=\uparrow, \downarrow$)
is the  electron occupation number operator of the impurity level,
and $\hat{c}^\dagger_s$ ($\hat{c}_s$) is the electron creation (annihilation) operator;
$\epsilon_s$ is the energy of the system level,
and $U$ is the electron-electron Coulomb interaction energy.
%

\begin{figure}[t]
  \centering
\includegraphics[width=0.8\columnwidth]{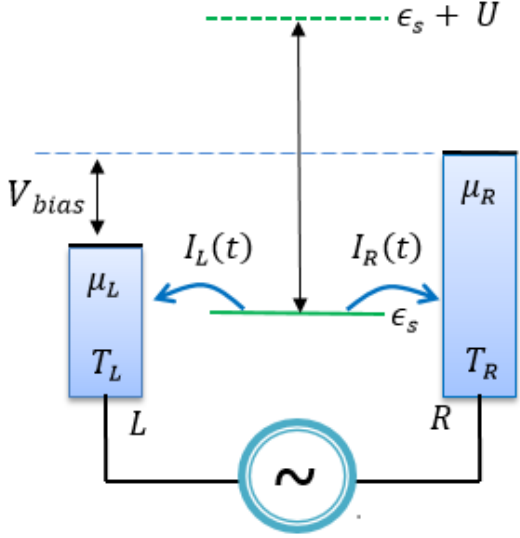}
\caption{Schematic diagram of a single-level impurity coupled with two electron reservoirs $(\alpha=L,R)$.
$\epsilon_s$ is the level energy for a spin-$s$ electron,
$U$ is the electron-electron Coulomb interaction energy,
and $\mu_\alpha$ and $T_\alpha$ are the chemical potential and temperature of the $\alpha$-reservoir.
$I_\alpha(t)$ represents the current flow between the impurity and the $\alpha$-reservoir
when a bias voltage $V_{\rm bias}$ is applied across the two reservoirs.}
\label{fig1}
\end{figure}

As shown in \Fig{fig1}, the impurity is coupled to two spin-unpolarized electron reservoirs ($\alpha = L, R$).
The Hamiltonian of the reservoirs (fermion bath) is
$\Hb=\sum_{\alpha k}\sum_{s}\epsilon_{\alpha k}\, \hat{d}^\dagger_{\alpha k s} \hat{d}_{\alpha k s}$,
where $\hat{d}^\dagger_{\alpha k s}$ ($\hat{d}_{\alpha k s}$) is the creation (annihilation) operator
of the $k$th level of the $\alpha$-reservoir.
The dissipative interaction between the impurity and the reservoirs is governed by
$\Hsb= \sum_\alpha \sum_s \hat{c}^\dagger_{s} \, \hat{F}_{\alpha s} + \hat{F}^\dagger_{\alpha s}\, \hat{c}_{s}$,
where $F_{\alpha s} = \sum_{k}t_{\alpha k} \hat{d}_{\alpha k s}$,
with $t_{\alpha k}$ being the coupling strength between the impurity level
and the $k$th level of $\alpha$-reservoir.
Both reservoirs have a spectral function of a Lorentz form, i.e.,
\begin{equation}
J_{\alpha} (\omega)\equiv \pi \sum_{k} |t_{\alpha k}|^2\delta(\omega - \epsilon_{\alpha k})
=\frac{\Gamma_\alpha}{2} \frac{W_\alpha^2}{(\omega-\Omega_\alpha)^2+W_\alpha^2},
\end{equation}
where $W_\alpha$ and $\Omega_\alpha$ are the band-width and band-center
of the $\alpha$th reservoir, respectively;
and $\Gamma_\alpha$  is the effective impurity-reservoir coupling strength.
In this work, the two reservoirs always have the same band-width
and band-center, i.e., $W_L = W_R = W$ and $\Omega_L = \Omega_R = \Omega$.

The AIM is truly an open system because $\Hsb$ allows the electrons to
transfer in and out of the impurity. The connecting electron reservoirs
may be at different temperatures ($T_R \neq T_L$), which could result in
the flow of thermal current.
Moreover, as shown in \Fig{fig1}, the applied bias voltage could change
the chemical potentials of the reservoirs, leading to the
electric current flow between the impurity and a reservoir.
We assume that at initial time $t=t_0$, the system and reservoirs
are decoupled and the isolated reservoirs are in thermal equilibrium states,
i.e., $\rhot(t_0)=\rhos(t_0)\,\rho^{\rm eq}_{_{\rm B}}$.
$\Hsb$ is turned on at $t=t_0^+$, which triggers the flow of electrons
between the impurity and the reservoirs.
During the time evolution of the composite system,
the quantities of primary interest
are the electron population on the impurity,
and the electric current flow into the reservoirs.

{\it Reservoir correlation functions}:
For non-interacting electron reservoirs which satisfy Gaussian statistics,
the two-time correlation functions, $C^\pm_\alpha(t,\tau)$,
account for the influence of the $\alpha$-reservoir on the impurity.
For reservoirs at thermal equilibrium state, the correlation
functions possess the translational symmetry in time, i.e.,
$\tilde{C}_\alpha^\sigma(t,\tau) = \tilde{C}_{\alpha}^\sigma(t-\tau)$,
and they are associated with the reservoir spectral function
via the fluctuation-dissipation theorem (FDT) as follows,\cite{jin2008exact}
\begin{equation} \label{correl}
\tilde{C}^\sigma_\alpha(t) = \frac{1}{2\pi}\int_{-\infty}^{\infty} d\omega\,
 e^{\sigma i \omega t} f^\sigma_\alpha(\omega) J^\sigma_\alpha (\omega).
\end{equation}
Here,
$f^\sigma_\alpha(\omega) =1 /[1+e^{\sigma \beta_\alpha (\omega - \mu_\alpha)}]$
is the Fermi function for electron $(\sigma=+)$ or hole $(\sigma=-)$,
$\beta_\alpha = 1/T_\alpha$  is the inverse temperature,
and $\mu_\alpha$ is the reservoir chemical potential (we set $\mu_\alpha^{\rm eq} = 0$).
For a single-level AIM, we have  $J^+_\alpha(\w) = J^-_\alpha(\w) = J_\alpha(\w)$.
The reservoir correlation functions expressed in \Eq{correl} satisfy the following relations\cite{jin2008exact}
\begin{equation}
[\tilde{C}^\sigma_\alpha(t)]^* = \tilde{C}^\sigma_\alpha(-t) =  e^{\sigma \beta_\alpha \mu_\alpha } \tilde{C}^{\bar{\sigma}}_\alpha (t-i\beta_\alpha ),
\end{equation}
where $\bar{\sigma}=-\sigma$. The first and second equalities
are the time reversal symmetry and the detailed balance relation, respectively.

By applying a time-dependent voltage $ V_\alpha(t)$, the electronic bands and the chemical potential
of the $\alpha$-reservoir undergo a homogeneous shift,
i.e., $\mu_\alpha(t) = \mu^{\rm eq}_\alpha  - e V_\alpha (t)$.
The reservoir correlation function thus includes an additional time-dependent phase factor,\cite{Zhe08184112}
\begin{equation} \label{corr}
C^\sigma_\alpha(t,\tau) = \exp \bigg[\sigma i \int_{\tau}^{t} dt'  \, V_\alpha(t')\bigg] \tilde{C}^\sigma_\alpha(t-\tau).
\end{equation}

{\it The MAS-SEOM}: By utilizing the MAS mapping of \Eq{map_eq1},
the stochastic reduced density matrix of a single-level impurity
coupled to two reservoirs is defined in the product space
$V\equiv V_{_{\rm S}} \otimes S_{1 \uparrow} \otimes S_{1\downarrow} \otimes S_{2 \uparrow} \otimes S_{2\downarrow}$
as\cite{han2019fermionic}
\begin{equation} \label{rho_s}
\trhos=\sum_{l_{1 \uparrow} \in S_{1 \uparrow}}\sum_{l_{1 \downarrow} \in S_{1 \downarrow}}\sum_{l_{2 \uparrow} \in S_{2\uparrow}} \sum_{l_{2\downarrow} \in S_{2 \downarrow}} \trhos ^{\left[l_{1 \uparrow}, l_{1 \downarrow} , l_{2 \uparrow}, \l_{2 \downarrow}\right]} ,
\end{equation}
where $V_{_{\rm S}}$ is the system subspace, and $S_{js}$ with $j=1,2$ and $s= \uparrow, \downarrow$
is the auxiliary space spanned by three pseudo-Fock-states, i.e., $S_{js} = \{-1,0,1 \}$.

The MAS-SEOM is written as\cite{han2019fermionic}
\begin{align} \label{Eq1}
\dtrhos  &= -i[\Hs, \trhos] + \lambda^{\frac{1}{2}}  \sum_{s=\uparrow,\downarrow}
 \sum_{\alpha=L,R} \Big[ \big(\hat{c}^\dagger_s \, Y_{1\alpha s} + Y_{2\alpha s}\, \hat{c}_s \big) \trhos  \nl
&\qquad \qquad \times e^{-\frac{i \pi}{4}}
+ e^{\frac{i \pi}{4}} \trhos \big(\hat{c}^\dagger_s\, Y_{3\alpha s} + Y_{4\alpha s}\, \hat{c}_s \big) \Big].
\end{align}
Here, the auxiliary operators $\{Y_{j\alpha s}; j=1,..,4\}$ are defined by
\begin{align} \label{parameters}
 Y_{1\alpha s} &\equiv v_{1\alpha st}\, X^-_{1s} + \tilde{g}^-_{\alpha st}\,, \nl
 Y_{2\alpha s} &\equiv v_{2\alpha st}\, X^+_{2s} - \tilde{g}^+_{\alpha st}\,,  \nl
 Y_{3\alpha s} &\equiv v_{3\alpha st}\, X^-_{1s} - i \tilde{g}^-_{\alpha st}\,, \nl
 Y_{4\alpha s} &\equiv v_{4\alpha st}\, X^+_{2s} + i \tilde{g}^+_{\alpha st}\,,
\end{align}
where $\{\tilde{g}_{\alpha s t}^\pm\}$ include the memory-convoluted noises
and the time-independent pseudo-operators $X^\pm_{js}$,
\begin{align} \label{def-tg-1}
  \tilde{g}^-_{\alpha s t} &=\lambda^{-1}\! \int_{t_0}^t\! d\tau \!
  \left\{[C^+_{\alpha}(t,\tau)]^\ast v_{4\alpha s\tau} -iC^-_{\alpha}(t,\tau)v_{2\alpha s\tau}\right\} X^-_{2s}, \nl
  \tilde{g}^+_{\alpha s t} &=\lambda^{-1}\!\int_{t_0}^t\! d\tau \!
  \left\{[C^-_{\alpha}(t,\tau)]^\ast v_{3\alpha s\tau} -iC^+_{\alpha}(t,\tau)v_{1\alpha s\tau}\right\} X^+_{1s}.
\end{align}
In \Eqs{parameters} and \eqref{def-tg-1}, $\{v_{j \alpha st}; j=1,\ldots,4\}$
are Gaussian white noises which satisfy
$\mathcal{M}(v_{j\alpha st})=0$ and $\mathcal{M}(v_{j \alpha st}v_{j' \alpha' s'\tau})=
\delta_{jj'}\delta_{\alpha \alpha'}\delta_{ss'}\delta(t-\tau)$,
where $\mathcal{M}$ denotes the stochastic average over all the random noises.
$\lambda$ is a reference energy which could take any positive value $(\lambda > 0)$.
The pseudo-operators $\{X^{\pm}_{js}; j=1,2\}$ can act on $\trhos$ from
both left and right; see Ref.~\onlinecite{han2019fermionic} for details.

The reduced density matrix of the impurity,
$\brhos = {\rm tr}_{_{\rm B}}(\rho_{_{\rm T}})$,
is obtained via the statistical average of
\begin{equation} \label{brhos}
\brhos = \la \trhos \ra \equiv \mathcal{M} \left( \trhos^{\left[0,0,0,0 \right]} \right) .
\end{equation}
The expectation value of any system operator $\hat{O}$ is thus
evaluated straightforwardly as
\begin{equation} \label{operator}
\langle \hat{O} \rangle_{_{\rm T}} =  {\rm tr}_{_{\rm S}} \left( \hat{O} \brhos  \right).
\end{equation}
In the above, $\rho_{_{\rm T}}$ is the density matrix of the total AIM,
and ${\rm tr}_{_{\rm S}}$ (${\rm tr}_{_{\rm B}}$) denotes the trace over
all the impurity (reservoir) degrees of freedom.

\subsection{Numerical representation of the system operators and the pseudo operators} \label{Numerical Representation}

As already been mentioned in \Sec{subsec:mas-seom},
the stochastic reduced density matrix $\trhos$ is defined in the product space $V$.
For simplicity and clarity, in the following we omit the spin index $s$, i.e.,
\begin{equation} \label{expan}
\trhos=\sum_{l_1 \in S_1}\sum_{l_2 \in S_2} \trhos^{[l_1, l_2]} .
\end{equation}
In the language of Grassmann algebra, the three pseudo-Fock-states of $S_j$,
$\{-1,0,1\}$, correspond to the time-independent g-numbers $\{ \eta_j,1,\bar{\eta}_j \}$.
In other words, $\trhos$ can be seen as a polynomial function of
the time-independent g-numbers $\{ \bar{\eta}_1, \eta_1, \bar{\eta}_2,\eta_2 \}$, i.e.,
\begin{equation}\label{poly}
\trhos= \sum_{p_1,p_2, p_3,p_4 \in \{0, 1\}} B_{p_1p_2p_3p_4} \,
\eta_1^{p_1} \, \bar{\eta}_1^{p_2}\, \eta_2^{p_3} \, \bar{\eta}_2^{p_4}.
\end{equation}
Here, all the monomials containing the conjugated pairs of g-numbers $\eta_j\bar{\eta}_j$
are suppressed or reduced to 1, i.e., $B_{p_1 p_2 p_3 p_4}=0$ if $p_1 = p_2 =1$ or $ p_3 = p_4 = 1$.
This is because a reduction procedure is found crucial to preserve
the even-order moments of time-dependent AGFs, and consequently
there is no pseudo-Fock-state in the auxiliary space $S_j$ that
corresponds to the conjugated pair $\eta_j\bar{\eta}_j$;
see Sec.~III of paper~I for details.

It should be emphasized that the order of g-numbers is of crucial importance.
By default all the monomials on the right-hand side of \Eq{poly} are in the normal order,
i.e., $\eta_j$ and $\bar{\eta}_j$ appear at the left of $\eta_{j'}$ and $\bar{\eta}_{j'}$
for $j < j'$, and $\eta_j$ is at the left of $\bar{\eta}_j$.
For instance, $\eta_1\eta_2$ and $\bar{\eta}_1\eta_2$ are considered to be in normal order,
while $\bar{\eta}_2\bar{\eta}_1$ and $\eta_2\eta_1$ are not.
All the monomials should be brought into normal order before any operator action.
The number of monomials in the expansion of \Eq{poly} increases with the size of the system.
For a system with $N_o$ orbitals (or levels) and $N_s$ spin directions,
the polynomial expansion comprises of $3^{\,2N_oN_s}$ terms.
For instance, the polynomial expansion for $\trhos$ of a spin-resolved
single-level AIM with $N_o=1$ and $N_s=2$, \Eq{poly} consists of 81 terms.

The actions of the pseudo-operators $X^\pm_{j}$ on $\trhos^{\left[l_1,l_2 \right]}$
are analogous to the actions of raising and lowering operators.
For instance, $X^{+}_1$ and $X^{+}_2$ raise the occupation number on
the pseudo-level in the auxiliary spaces $S_1$ and $S_2$, respectively.
Similarly, $X^{-}_1$ and $X^{-}_2$ lower the occupation numbers of
the pseudo-levels 1 and 2, respectively. Specifically, we have
\begin{align} \label{Eq.3}
\begin{split}
&X^+_1\trhos^{[l_1,l_2]} = (-1)^{l_1+l_2}\trhos^{[l_1,l_2]}X^+_1 = \chi^{l_1}_{\{-1,0\}}(-1)^{l_1} \trhos^{[l_1+1,l_2]} \ , \\ &
X^-_1 \trhos^{[l_1,l_2]} = (-1)^{l_1+l_2}\trhos^{[l_1,l_2]}X^-_1 = \chi^{l_1}_{\{0,1\}} \trhos^{[l_1-1,l_2]}  \ ,\\ &
X^+_2\trhos^{[l_1,l_2]} = (-1)^{l_1+l_2}\trhos^{[l_1,l_2]}X^+_2 = \chi^{l_2}_{\{-1,0\}}(-1)^{l_1+l_2} \trhos^{[l_1,l_2+1]} \ , \\&
X^-_2{\tilde{\rho}}_\text{S}^{[l_1,l_2]} = (-1)^{l_1+l_2}\trhos^{[l_1,l_2]}X^-_2 = \chi^{l_2}_{\{0,1\}}(-1)^{l_1}\trhos^{[l_1,l_2-1]}.
\end{split}
\end{align}
Here, $\chi$ is a step function, i.e., $\chi^{l_j}_\mathbb{Z} = 1$
if ($l_j \in \mathbb{Z}$) or 0 (if $l_j \notin \mathbb{Z})$,
which restricts the result of action of $X^\pm_j$ within the auxiliary space $S_j$.
The prefactors in \Eq{Eq.3} are $1$ or $-1$, which tell us how many swaps
are required to rearrange and bring the g-numbers into normal order.

In a very concise form, the action of $\{X^\pm_{j}; j=1,2\}$ on $\trhos^{[l_1,l_2]}$ can be expressed as
\begin{equation} \label{action}
\begin{matrix}
& &  &  \xrightarrow{\textstyle X^+_2 \trhos }  & &\\& \trhos^{[-1,-1]} & \xleftrightarrow[-1]{1} & \trhos^{[-1,0]} & \xleftrightarrow[-1]{-1} & \trhos^{[-1,1]} &\\ & \scriptstyle{-1} \displaystyle \updownarrow \, \scriptstyle{1} & & \scriptstyle{-1} \displaystyle \updownarrow \, \scriptstyle{1} & &\scriptstyle{-1} \displaystyle \updownarrow \, \scriptstyle{1} \\  \textstyle{X^+_1 \trhos} \Big\downarrow & \trhos^{[0,-1]} & \xleftrightarrow[1]{-1} & \trhos^{[0,0]} & \xleftrightarrow[1]{1} & \trhos^{[0,1]} & \Big \uparrow \textstyle{X^-_1 \trhos } \\ & \scriptstyle{1} \displaystyle \updownarrow \, \scriptstyle{1} & & \scriptstyle{1} \displaystyle \updownarrow \, \scriptstyle{1} & & \scriptstyle{1} \displaystyle \updownarrow \, \scriptstyle{1} \\ & \trhos^{[1,-1]} & \xleftrightarrow[-1]{1} & \trhos^{[1,0]} & \xleftrightarrow[-1]{-1} & \trhos^{[1,1]} & \\ & & & \xleftarrow[\textstyle X^-_2 \trhos ]  & &
\end{matrix}
\end{equation}
Here, the factors ($-1$ or 1) to the left and right of $\updownarrow$
are associated with the actions $X^+_1 \trhos$ and $X^-_1 \trhos$, respectively;
while the factors ($-1$ or 1) above and below the arrows $\leftrightarrow$
are associated with the actions $X^+_2 \trhos$ and $X^-_2 \trhos$, respectively.

Based on the correspondence between the pseudo-Fock-states and the g-numbers,
the actions of $X^\pm_j$ on $\trhos^{[l_1,l_2]}$ can be translated into
the language of Grassmann algebra as follows,
\begin{equation} \label{grass-action}
\begin{matrix}
& &  &  \xrightarrow{\textstyle \hat{r}(\bar{\eta}_2 \trhos) }  & & \\&   \eta_1\eta_2 &  \xleftrightarrow[-1]{1}  &   \eta_1 & \xleftrightarrow [-1]{-1} &  \eta_1\bar{\eta}_2  &\\ & \scriptstyle{-1} \displaystyle \updownarrow \, \scriptstyle{1}  & & \scriptstyle{-1} \displaystyle \updownarrow \, \scriptstyle{1}   & & \scriptstyle{-1} \displaystyle \updownarrow \, \scriptstyle{1}    \\  \textstyle{\hat{r}(\bar{\eta}_1  \trhos)}  \Big\downarrow &  \eta_2 &   \xleftrightarrow[1]{-1} &   1 &  \xleftrightarrow[1]{1} & \bar{\eta}_2 & \Big \uparrow \textstyle{ \hat{r}(\eta_1 \trhos)} \\ &  \scriptstyle{1} \displaystyle \updownarrow \, \scriptstyle{1} & &  \scriptstyle{1} \displaystyle \updownarrow \, \scriptstyle{1} & & \scriptstyle{1} \displaystyle \updownarrow \, \scriptstyle{1} \\ &  \bar{\eta}_1\eta_2 & \xleftrightarrow[-1]{1} & \bar{\eta}_1 & \xleftrightarrow [-1]{-1} &  \bar{\eta}_1  \bar{\eta}_2 & \\ & & &  \xleftarrow[\textstyle \hat{r} (\eta_2 \trhos)]  & &
\end{matrix}
\end{equation}
Here, $\hat{r}$ is a linear operator introduced for describing
the reduction procedure for the product of g-numbers $\{ \eta_j, \bar{\eta}_j \}$;
see Sec.~III of Paper~I.
\begin{equation}
\hat{r}(1)=1 , \ \hat{r}(\eta_j)=\eta_j , \ \hat{r}(\bar{\eta}_j)=\bar{\eta}_j , \  \hat{r}(\eta_j \bar{\eta}_j)=1 .
\end{equation}
In \Eq{grass-action}, the factors  ($-1$ or 1) to the left and right of $\updownarrow$
are associated with the operations $\hat{r}(\bar{\eta}_1 \trhos)$ and $\hat{r}({\eta}_1 \trhos)$, respectively;
while the factors ($-1$ or 1) over and below the arrows $\leftrightarrow$
are associated with the operations $\hat{r}(\bar{\eta}_2 \trhos)$ and $\hat{r}({\eta}_2 \trhos)$, respectively;

By comparing \Eq{grass-action} with \Eq{action}, we can establish a
one-to-one correspondence between the action of $X^\pm_j$ on pseudo-Fock-states
and the action of $\hat{r}$ on the monomials of g-numbers.
Specifically, the action of $X^{+}_j$ on $\trhos$ corresponds to the operation $\hat{r}(\bar{\eta}_j \trhos)$,
while the action of $X^{-}_j$ on $\trhos$ corresponds to the operation $\hat{r}(\eta_{j} \trhos)$.
For instance, the pseudo-Fock-state $|\!-\!1,1\ra$ corresponds to the normal-ordered monomial
of g-numbers $\eta_1\bar{\eta}_2$. The action of $X^+_1$ from the left side will change it to a new state
$-|0,1\ra$, which corresponds to $-\bar{\eta}_2$.
If the reduction operator is used, we have
$\hat{r}(\bar{\eta}_1\eta_1 \bar{\eta}_2) = - \hat{r}(\eta_1\bar{\eta}_1) \hat{r}(\bar{\eta}_2) = -\bar{\eta}_2$.
Here, the original pair of g-numbers  $\eta_1\bar{\eta}_1$ is reduced to 1 by $\hat{r}$,
where the factor of $-1$ arises from rearranging the g-numbers into normal order.

Now let's come to the numerical representation of the system operators.
As we know that with second quantization formulation,
the system Hamiltonian can be written in terms of creation and annihilation operators of the system.
For an impurity of $N_o$ levels and $N_s$ spin directions,
the dimension of the system Hilbert space is $(2N_s)^{N_o}$.
We thus need to use matrices of the size $(2N_s)^{N_o} \times (2N_s)^{N_o}$
to represent the operators $\{\hat{c}_{\nu s}; \nu=1,\ldots,N_o\}$.
For instance, for a spinless single-level AIM ($N_s=1, N_o=1$), only two Fock-states,
$|0\rangle$ (vacant) and $|1\rangle$ (occupied), span the Hilbert space of the impurity.
and the annihilation operator $\hat{c}$ is represented by a $2\times 2$ matrix.
\begin{equation}
\hat{c} =|0\rangle\langle 1| =
\begin{pmatrix}
0 & 1 \\ 0 & 0
\end{pmatrix} .
\end{equation}
For a spin-resolved single-level AIM ($N_s=2, N_o=1$), four Fock-states,
$|0\rangle$ (vacant), $|\!\! \uparrow\ra$ (singly occupied by a spin-up electron),
$|\! \downarrow \ra$ (singly occupied by a spin-down electron) and $|2\rangle$ (doubly occupied),
span the Hilbert space of the impurity.
In this case, the creation and annihilation operators are represented by $4 \times 4$ matrices.
\begin{align}
\hat{c}_\uparrow &=|0\rangle\langle \uparrow\!| + |\!\downarrow\rangle\langle 2| =
\begin{pmatrix}
0 & 1 & 0 & 0 \\ 0 & 0 & 0 & 0 \\ 0 & 0 & 0 & 1 \\ 0 & 0 & 0 & 0
\end{pmatrix} ,  \nl
\hat{c}_\downarrow &=|0\rangle\langle \downarrow\!| - |\!\uparrow \rangle\langle 2| =
\begin{pmatrix}
0 & 0 & 1 & 0 \\ 0 & 0 & 0 & -1 \\ 0 & 0 & 0 & 0 \\ 0 & 0 & 0 & 0
\end{pmatrix} .
\end{align}
The number of nonzero elements (1 or $-1$) for any creation or annihilation operator is $(2N_s)^{N_o} /2$.
The matrix representation of the creation and annihilation operators is not unique.
Any set of matrices satisfying the anti-commutation relations,
$\{\hat{c}_{\nu s}, \hat{c}_{\nu's'}\}=\{\hat{c}^\dagger_{\nu s}, \hat{c}^\dagger_{\nu's'}\}=0$
and $ \{\hat{c}_{\nu s}, \hat{c}^\dagger_{\nu's'}\}=\delta_{\nu\nu'}\delta_{ss'}$, can be used.

It is also worth emphasizing that the system creation and annihilation operators
act on every component of the stochastic reduced density matrix $\trhos$.
For instance, the action of $\hat{c}$ on $\trhos$ in the product space
is carried out as follows.
\be
\hat{c}\,\trhos= \sum_{l_1\in S_1} \sum_{l_2 \in S_2} \hat{c}\,\trhos^{[l_1, l_2]}.
\ee

\section{Calculation of electric current between system and reservoir} \label{sec:current}
\subsection{Analytic expression of electric current in the rigorous SEOM formalism}

As already mentioned, \Eq{operator} can be used to calculate the expectation value
of any system operator. However, the particle current operator is not a pure system operator,
as it involves both the system and reservoir's degrees of freedom.
Thus, we need to derive the analytic expression for the expectation value of
current operator from the total density matrix $\rhot$.

Consider a spinless single-level AIM in which the impurity level
is coupled to two reservoirs $(\alpha = L, R)$; see \Fig{fig1}.
The time-dependent electric current flowing into the $\alpha$-reservoir is given by\cite{jin2008exact}
\begin{align} \label{curr}
I_\alpha (t)=-\frac{d}{dt}\langle \hat{N}_{\alpha} \rangle _{_{\rm T}}
 &= -{\rm tr}_{_{\rm T}} \left(\hat{N}_{\alpha}\, \dot{\rho}_{_{\rm T}}(t) \right) \nl
 & = i\,{\rm tr}_{_{\rm T}}\left(\big[ \hat{N}_{\alpha}, \Hsb \big] \rhot \right)  \nl
 &= i\, {\rm tr}_{_{\rm T}} \left(  \hat{c}\, \rhot\, \hat{F}^\dagger_\alpha
   -  \hat{F}_\alpha\, \rhot\,\hat{c}^\dagger  \right),
\end{align}
where $\hat{N}_{\alpha}$ is the electron number operator of the $\alpha$-reservoir.

In Sec.~II\,C of paper~I, we have demonstrated that the dynamics of
the system and reservoir (bath) can be decoupled by introducing the time-dependent AGFs.
This leads to the formally exact SEOM for the system reduced density matrix $\rhos$ and
that for the bath density matrix $\rhob$. The total density matrix is exactly
recovered as $\rhot=\left\langle \rhos \rhob \right\rangle = \left\langle \rhob \rhos \right\rangle$,
where $\la \cdots \ra$ denotes the stochastic average over all the AGFs.
Accordingly, the electric current is expressed as
%
\begin{align} \label{deriv}
I_\alpha(t) &=i\, {\rm tr}_{_{\rm T}}
\big( \big\langle \hat{c}\, \rhos\rhob \hat{F}^\dagger_\alpha - \hat{F}_\alpha\, \rhob \rhos \hat{c}^\dagger \big\rangle \big) \nl
&=i \,\big\langle \, {\rm tr}_{_{\rm S}} (\hat{c}\rhos)\, {\rm tr}_{_{\rm B}} \big(  \rhob \hat{F}^\dagger_\alpha \big)
 - {\rm tr}_{_{\rm B}} \big( \hat{F}_\alpha \,\rhob \big)\,{\rm tr}_{_{\rm S}} (\rhos \hat{c}^\dagger)\, \big\rangle  \nl
&= i\, \big\langle \, {\rm tr}_{_{\rm S}}(\hat{c} \trhos)\, \tilde{F}^\dagger_{\alpha} \,
 - \bar{F}_{\alpha} \, {\rm tr}_{_{\rm S}} ( \trhos \hat{c}^\dagger ) \, \big\rangle.
\end{align}
where ${\rm tr}_{_{\rm B}}$ and ${\rm tr}_{_{\rm S}}$ represent
the trace over the reservoir and system's degrees of freedom, respectively;
and we have used the definitions
\be \label{F-define}
\trhos \equiv \rho_{_{\rm S}} {\rm tr}_{_{\rm B}}(\rhob), \,
\tilde{F}^\dagger_{\alpha} \equiv \frac{{\rm tr}_{_{\rm B}}\big( \rhob \hat{F}^\dagger_{\alpha}\big)}{{\rm tr}_{_{\rm B}}( \rhob )} ,  \,  \bar{F}_{\alpha}\equiv \frac{{\rm tr}_{_{\rm B}} \big(\hat{F}_\alpha \, \rhob\big)}{{\rm tr}_{_{\rm B}}(\rhob)} .
\ee
The bath density matrix $\rhob$ satisfies the SEOM of\cite{han2019fermionic}
\begin{align} \label{eom-rhob-1}
  \drhob = & -i[\Hb, \rhob] + \lambda^{-\frac{1}{2}}
  \sum_{\alpha=L,R} \big[ e^{-\frac{i\pi}{4}}  \big( \bar{\eta}_{1\alpha t} \hat{F}_\alpha + \hat{F}^\dagger_\alpha \eta_{2\alpha t}\big) \rhob \nl
  & + e^{\frac{i\pi}{4}} \rhob \big( \bar{\eta}_{3\alpha t} \hat{F}_\alpha + \hat{F}^\dagger_\alpha \eta_{4\alpha t}\big) \,\big].
\end{align}
Taking the trace over all the bath's degrees of freedom for both sides of \Eq{eom-rhob-1}, we have
\begin{align} \label{rhob-1}
  \frac{d}{dt}{\rm tr}_{_{\rm B}} ( \rhob ) &= \lambda^{-\frac{1}{2}} \! \sum_{\alpha=L,R}
  \big[\, e^{-\frac{i\pi}{4}} \left( \bar{\eta}_{1\alpha t} \bar{F}_\alpha- \eta_{2\alpha t} \bar{F}^\dagger_\alpha  \right) + e^{\frac{i\pi}{4}} \nl
  & \qquad \times \big(\! -\tilde{F}_\alpha \,\bar{\eta}_{3\alpha t} + \tilde{F}^\dagger_\alpha \,\eta_{4\alpha t} \big)\, \big] \,
  {\rm tr}_{_{\rm B}}(\rhob) .
\end{align}
Here, we have defined
$\tilde{F}_\alpha \equiv {\rm tr}_{_{\rm B}} ( \rhob \hat{F}_\alpha) /{\rm tr}_{_{\rm B}}( \rhob )$
and $\bar{F}^\dagger_\alpha \equiv {\rm tr}_{_{\rm B}} ( \hat{F}^\dagger_\alpha\, \rhob ) / {\rm tr}_{_{\rm B}}( \rhob )$,
and  $\bar{F}_\alpha$ and $\tilde{F}^\dagger_\alpha$ are defined in \Eq{F-define}.
In \Eq{rhob-1}, we have used the important property that the AGFs anti-commute with
the reservoir operators $\hat{F}^\dagger_\alpha$ and $\hat{F}_\alpha$.
After some rearrangement of terms, \Eq{rhob-1} becomes
\begin{align} \label{tr-rhob-0}
 \frac{d}{dt} \ln \left[ \text{tr}_{_{\rm B}} \left( \rhob \right)\right] &= \lambda^{-\frac{1}{2}} e^{-\frac{i\pi}{4}}
 \! \sum_{\alpha=L,R} \big[\,\big(  \bar{\eta}_{1\alpha t} \bar{F}_\alpha -i \tilde{F}_\alpha \,\bar{\eta}_{3\alpha t} \big) \nl
& \quad - \big( \eta_{2\alpha t} \bar{F}^\dagger_\alpha -  i\tilde{F}^\dagger_\alpha \,\eta_{4\alpha t}  \big) \big] .
\end{align}
On the other hand, because the non-interacting electron reservoirs satisfy the
Gaussian statistics, the formal solution of \Eq{eom-rhob-1} can be obtained by
utilizing the Magnus expansion\cite{tannor2007introduction} in the
$\Hb$-interaction picture as follows,\cite{han2019fermionic}
\begin{align}   \label{trace-rhob}
  {\rm tr}_{_{\rm B}} ( \rhob ) &= \exp \bigg\{ \sum_{\alpha = L,R}\, \int_{t_0}^t  d\tau
   \Big[ \left(\bar{\eta}_{1\alpha \tau} - i\bar{\eta}_{3\alpha \tau} \right) g^{-}_{\alpha \tau} \nl
   & \qquad \qquad + \left(\eta_{2\alpha \tau} - i\eta_{4\alpha \tau} \right)  g^{+}_{\alpha \tau} \Big] \bigg\} ,
\end{align}
with
\begin{align}
  g^-_{\alpha t} &=  \lambda^{-1} \int_{t_0}^t d\tau \,
  \big\{ [C^+_\alpha(t,\tau)]^\ast \eta_{4\alpha \tau} - i C^-_\alpha(t,\tau)\eta_{2\alpha \tau} \big\} , \nl
  g^+_{\alpha t} &=  \lambda^{-1} \int_{t_0}^t d\tau \,
  \big\{ [C^-_\alpha(t,\tau)]^\ast \bar{\eta}_{3\alpha \tau} - iC^+_\alpha(t,\tau)\bar{\eta}_{1\alpha \tau} \big\}. \label{gbar-1}
\end{align}
To have a direct comparison with \Eq{tr-rhob-0}, take the time derivative of
both sides of \Eq{trace-rhob}:
\be \label{d-trace-rhob}
\frac{d}{dt} \ln \left[ {\rm tr}_{_{\rm B}} ( \rhob)\right] =
\left(\bar{\eta}_{1\alpha t} - i\bar{\eta}_{3\alpha t} \right) g^-_{\alpha t}
+ \left(\eta_{2\alpha t}  - i\eta_{4\alpha t} \right) g^+_{ \alpha t}.
\ee
By comparing \Eq{tr-rhob-0} and \Eq{d-trace-rhob}, we get
$ g^-_{\alpha t} = \lambda^{-\frac{1}{2}} e^{-\frac{i\pi}{4}}\bar{F}_\alpha$,
$ g^+_{\alpha t} = -\lambda^{-\frac{1}{2}} e^{-\frac{i\pi}{4}} \bar{F}^\dagger_\alpha$,
$ g^-_{\alpha t} = -\lambda^{-\frac{1}{2}} e^{-\frac{i\pi}{4}}\tilde{F}_\alpha$,
and $ g^+_{\alpha t} = \lambda^{-\frac{1}{2}} e^{-\frac{i\pi}{4}} \tilde{F}^\dagger_\alpha$,
which can be rewritten as
\begin{equation} \label{Fg}
\begin{aligned}
\bar{F}^\dagger_\alpha&=-\lambda^\frac{1}{2}e^{\frac{i \pi}{4}} g^+_{\alpha t} \ , \quad  &\bar{F}_\alpha &=\lambda^\frac{1}{2}e^{\frac{i \pi}{4}}   g^-_{\alpha t}  \ , \\
\tilde{F}^\dagger_\alpha&=\lambda^\frac{1}{2}e^{\frac{i \pi}{4}} g^+_{\alpha t} \ , \quad  &\tilde{F}_\alpha &=-\lambda^\frac{1}{2}e^{\frac{i \pi}{4}}   g^-_{\alpha t} \ .
\end{aligned}
\end{equation}

\subsection{Computation of electric current in the MAS-SEOM method}

By inserting \Eq{Fg} into \Eq{deriv} and utilizing the MAS mapping scheme
with $X_{1}^\pm = X_{3}^\pm$ and $X_{2}^\pm = X_{4}^\pm$, we obtain
the following analytic expression of electric current,
\be \label{current_1}
I_{\alpha}(t) = \lambda^\frac{1}{2}\, e^{\frac{-i \pi}{4}}\, \big\langle\,  \tilde{g}^+_{\alpha t}\, {\rm tr}_{_{\rm S}} ( \hat{c} \trhos)
+ \tilde{g}^-_{\alpha t} \, {\rm tr}_{_{\rm S}} ( \trhos  \hat{c}^\dagger )  \,\big\rangle,
\ee
where
\begin{align} \label{gw}
\tilde{g}^-_{\alpha t}  &= \lambda^{-1} (w_{4 \alpha  t}  - i w_{2 \alpha t})\, X^-_2,  \nl
\tilde{g}^+_{\alpha t}  &= \lambda^{-1} (w_{3 \alpha  t}  - i w_{1 \alpha t})\, X^+_1,
\end{align}
with
\begin{align} \label{eq_omega}
w_{1\alpha  t} &=  \int_{t_0}^t \, C^+_{\alpha }(t,\tau)\, v_{1 \alpha \tau} \, d\tau , \nl
w_{2 \alpha  t} & = \int_{t_0}^t \, C^-_{\alpha } (t,\tau) \, v_{2 \alpha \tau}\, d\tau , \nl
w_{3 \alpha  t} &= \int_{t_0}^t \, [C^-_{\alpha } (t,\tau)]^*\, v_{3 \alpha \tau}\, d\tau, \nl
w_{4 \alpha  t} &= \int_{t_0}^t \, [C^+_{\alpha }(t,\tau)]^*\, v_{4 \alpha \tau}\, d\tau.
\end{align}
Here, the memory-convoluted noises $\{w_{j\alpha t}\}$ can be generated
by employing the fast Fourier transform technique.\cite{shao2010rigorous}
After some rearrangement, \Eq{current_1} is recast into the form of
\begin{align} \label{current}
I_\alpha (t) &= \lambda^{-\frac{1}{2}}e^{-\frac{i\pi}{4}} \,
\big\langle\, (w_{3\alpha  t}  - iw_{1\alpha  t})\, {\rm tr}_{_{\rm S}} (X^+_{1} \hat{c} \trhos)  \nl
& \qquad \quad + (w_{4\alpha  t}  - iw_{2\alpha  t})\, {\rm tr}_{_{\rm S}} ( X^-_{2} \trhos  \hat{c}^\dagger)\, \big\rangle.
\end{align}
Here, the average $\la \cdots \ra$ is defined by \Eq{brhos}.

If the impurity is coupled to only one reservoir, the electrons entering into the
reservoir come exclusively from the impurity. Thus, the electric current can be obtained
from the conservation of particles, i.e.,
\begin{equation}\label{sys_curr}
I(t)=\frac{d}{dt}\langle \hat{n} \rangle_{_{\rm T}} =
{\rm tr}_{_{\rm S}} \big( \hat{c}^\dagger \hat{c} \, \langle \dtrhos \rangle \big).
\end{equation}

\section{Asymptotic behavior of memory-convoluted noises at ultra-low temperatures} \label{sec:stability}

Numerical stability of the MAS-SEOM depends critically on the amplitudes of the
involving instantaneous and memory-convoluted noises,  $\{v_{jt}\}$ and $\{w_{jt}\}$,
which drive the reduced system dynamics.
In particular, it is important that the amplitudes of the color noises $\{w_{jt}\}$
do not grow with time in the asymptotic limit.
For simplicity, consider
\begin{equation} \label{omega_3}
w_{t}= \int_{0}^{t} C(t-\tau)\,v_\tau\,  d\tau,
\end{equation}
where we have set $t_0 = 0$.
$C(t-\tau)$ is the reservoir correlation function, and $\{v_\tau\}$ are Gaussian white noises
which satisfy $ \mathcal{M} \left( v_t v_\tau \right) = \delta(t-\tau)$.
By the definition of \Eq{omega_3}, $w_t$ has the dimension of $t^{\frac{1}{2}}$, same as a Wiener process.

In the MAS-SEOM of \Eq{Eq1}, the contributions of $\{v_{jt}\}$ and $\{w_{jt}\}$
to $\trhos$ are scaled by $\lambda^{\frac{1}{2}}$ and $\lambda^{-\frac{1}{2}}$, respectively.
Therefore, one could adjust the value of $\lambda$
to balance the amplitudes of the instantaneous and memory-convoluted noises
and thus optimize the numerical performance.
Enlarging the value of $\lambda$ will amplify the instantaneous noises
while reduce the amplitudes of the memory-convoluted noises, and vice versa.
Nevertheless, any $\lambda$ should yield the same $\la\trhos\ra$.

Now let us examine the intrinsic behavior of $w_t$.
If $C(t)$ is a non-decaying function such as a constant, i.e., $C(t) = C_0$
(see Appendix~\ref{append:two_level} for a closed two-level system),
the auto-correlation of $w_t$ will be
\begin{equation} \label{auto-w2-1}
\mathcal{M} ( w^2 ) =C^2_0 \int_{0}^{t} d\tau_1 \int_{0}^{t} d\tau_2 \,
\mathcal{M}\left( v_{\tau_1} v_{\tau_2} \right) = C^2_0\,t.
\end{equation}
Obviously, the amplitude of $w_t$ keeps increasing with time.
Consequently, the MAS-SEOM becomes unstable soon after a certain period of time.
Such kind of asymptotic instability originates from the much too
strong environmental fluctuations, and is hard to avoid within
the stochastic framework.

For an open quantum system, $C(t)$ always decays with time.
In the long time limit, $C(t \rightarrow \infty) = 0 $.
Consider an electron reservoir with a finite band-width $W$,
the reservoir spectral function is
$J(\omega)=2\pi \Gamma$ for $|\omega|< W$ and $J(\omega)=0$ for $|\omega|> W$.
At zero temperature, the reservoir correlation function is obtained
through the FDT of \Eq{correl} as follows,
\begin{equation} \label{ct-zero-temp-1}
C(t)=\int_{-W}^{0} e^{-i\omega t}\, \Gamma\, d\omega = \frac{i\Gamma}{t} (1-e^{iWt}).
\end{equation}
The auto-correlation of $w_t$ is
\begin{align}  \label{auto-wt-1}
\mathcal{M} \big(w_t^2 \big) &= \int_{0}^{t}d\tau_1 \int_{0}^{t} d\tau_2 \,
 C(t-\tau_1)C(t-\tau_2) \mathcal{M} \left( v_{\tau_1} v_{\tau_2} \right) \nl
& =4\Gamma^2 \int_{0}^{t} e^{iW\tau} \Big( \frac{\sin(W\tau/2)}{\tau}  \Big)^2 \, d\tau.
\end{align}
The squared term on the right-hand side is real and non-negative, so we have
\begin{align}
\mathcal{M}\big(w_t^2 \big) & \leq 4\Gamma^2 \int_{0}^{t} \Big ( \frac{\sin(W\tau/2)}{\tau}  \Big )^2 d\tau \nl
 &= 4\Gamma^2 \bigg (\frac{W\, {\rm Si}(Wt)}{2} -\frac{\sin^2(Wt/2)}{t} \bigg ) \nl
 & < 2\Gamma^2 W \,{\rm Si}(Wt) < 4\Gamma^2 W.
\end{align}
Here, ${\rm Si}(x) \equiv \int_{0}^{x} \frac{\sin(u)}{u} du$ is the
sine integral which has an upper bound, and ${\rm Si}(\infty)=\pi/2$.
Therefore, the amplitude of $w_t$ does not grow with time even for
the slowly decaying function $C(t)$ at $T=0$.
This affirms the MAS-SEOM is asymptotically stable for general
open quantum systems at ultra-low temperatures.

\section{Results and Discussions} \label{sec:results}


\subsection{Accuracy and efficiency of MAS-SEOM} \label{subsec:accuracy}

In order to demonstrate the accuracy and applicability of the MAS-SEOM method,
we present some benchmark results for the single-level AIM.
We assume that the impurity and reservoirs are initially decoupled, i.e.,
$\rhot(t_0)=\rho_0 \rho^{\rm eq}_{_{\rm B}}$.
The impurity-reservoir couplings are turned on at $t = 0^{+}$,
which triggers the dissipative dynamics.
We consider two scenarios: (i) the dissipative relaxation process
brings the impurity and reservoirs towards a global thermal equilibrium state;
and (ii) the dissipative dynamics driven by the applied voltage
keeps the whole system in a non-equilibrium state.

The stochastic calculations are carried out as follows.
The time evolution of $\trhos$ is obtained by propagating the
MAS-SEOM of \Eq{Eq1} by employing certain stochastic integrators. 
The interested key quantities, such as the electron occupation number
on the impurity level, $n_s = \la \hat{n}_s \ra$,
and the electric current flow into the $\alpha$-reservoir, $I_\alpha(t)$,
are computed by \Eq{operator} and \Eq{current}, respectively.

In the following, we adopt the units suitable for quantum dot systems,
i.e., the energies are in units of meV, and the units for time and electric current
are ps and pA, respectively. These units can be easily rescaled to
be used for molecular systems. For the latter, the corresponding units
for energy, time and current are eV, fs and nA, respectively.

\begin{figure}[t]
\centering
\includegraphics[width=\columnwidth]{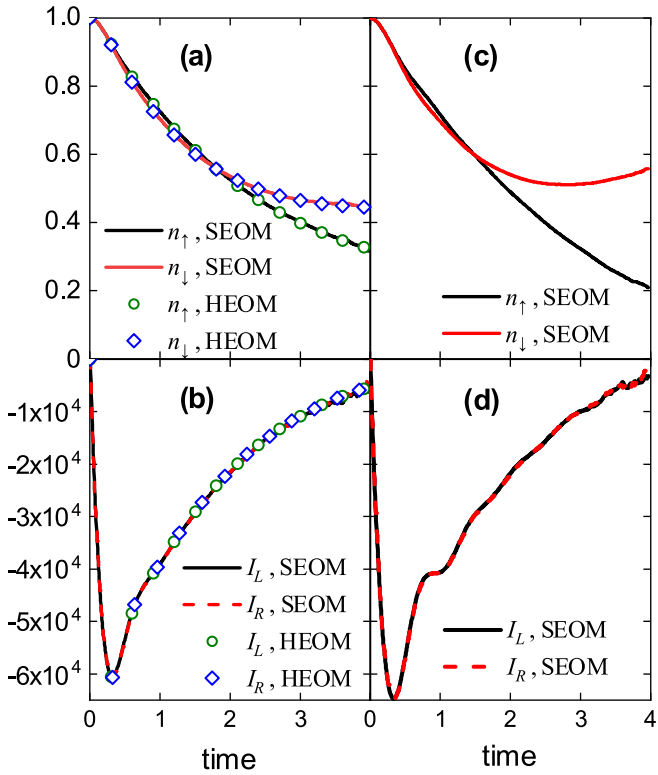}
\caption{Evolution of (a) $n_s$ and (b) $I_\alpha$
during the relaxation dynamics of a single-level AIM at a high temperature $T_L=T_R=1.0$,
and the evolution of (c) $n_s$ and (d) $I_\alpha$ of the same AIM
at an ultra-low temperature $T_L=T_R=1.0 \times 10^{-5}$.
The impurity level is initially doubly occupied, i.e., all the elements of $\rho_0$
are zero except $(\rho_0)_{44}=1$. The parameters of the AIM are:
$\epsilon_{\uparrow}=0.5$, $\epsilon_{\downarrow}=-0.5$,
$U=5.0$, $\Gamma_L=\Gamma_R=0.25$, $\Omega=0$, and $W=5.0$;
see the main text in \Sec{subsec:accuracy} for the description of units.
The Euler-Maruyama algorithm\cite{Klo92} is employed for the propagation of the MAS-SEOM,
with the time step $dt=0.005$, $\lambda=1.0$, and the number of trajectories
$N_{\rm traj}=3.5 \times 10^7$.
In (a) and (b) the results of the full HEOM are also displayed as a reference for comparison.}
\label{fig2}
\end{figure}

\begin{figure}[t]
\centering
\includegraphics[width=\columnwidth]{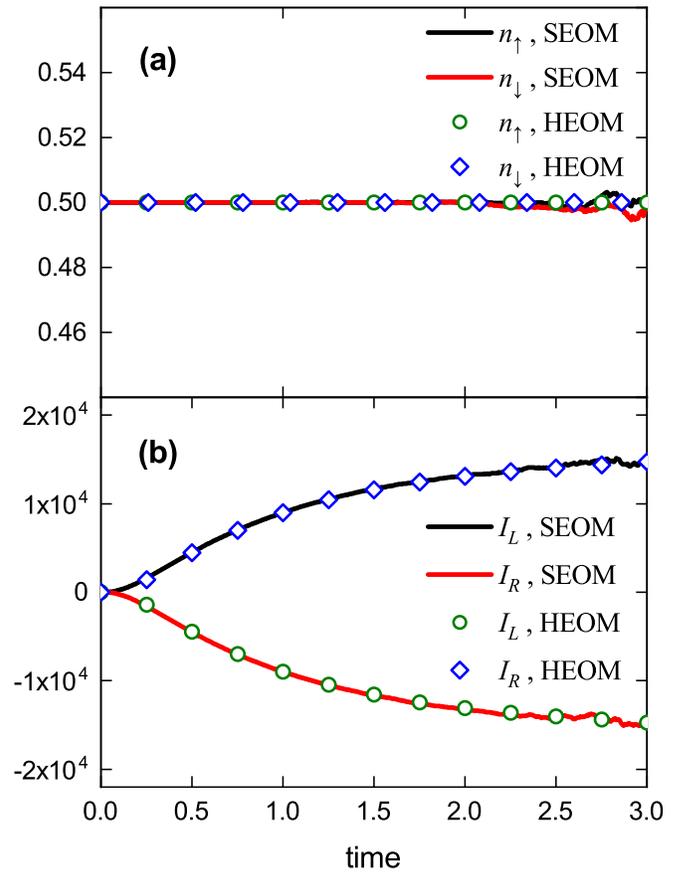}
\caption{Evolution of (a) $n_s$ and (b) $I_\alpha$ driven by a constant
bias voltage $V_L=-V_R=0.2$ for a single-level AIM at a low temperature $T_L=T_R=0.01$.
Initially, the impurity level has equal probabilities for all the four Fock-states,
i.e., $(\rho_0)_{ij}=\frac{1}{4}\delta_{ij} \, \{i,j=1,\ldots,4 \}$.
The parameters of the AIM are: $\epsilon_{\uparrow}=\epsilon_{\downarrow}=-0.5$,
$U=1.0$, $\Gamma_L=\Gamma_R=0.25$, $\Omega=0$, and $W=5.0$;
see the main text in \Sec{subsec:accuracy} for the description of units.
The Euler-Maruyama algorithm is employed for the propagation of the MAS-SEOM,
with $dt=0.005$, $\lambda=1.0$, and $N_{\rm traj}=7 \times 10^7$.
The results of the full HEOM are also displayed as a reference for comparison.}
\label{fig3}
\end{figure}

\textit{Relaxation dynamics}:
We consider the situation that the impurity level is initially doubly occupied,
i.e., $(\rho_0)_{44} = 1$ and all the other elements of $\rho_0$ are zero.
The time evolution of $n_s(t)$ and $I_\alpha(t)$ are computed and depicted in \Fig{fig2}
for both a high and a low temperatures.
For the former, the HEOM method implemented in the HEOM-QUICK program\cite{ye2016heom}
is also employed, and the results are displayed in \Fig{fig2}
for a direction comparison with those of the MAS-SEOM.


From \Fig{fig2}, it is clear that, while the relaxation dynamics of the AIM
at the high and low temperatures are overall similar, the low-temperature
dynamics exhibits some quantum oscillation features in the transient regime.
This indicates that the non-Markovian memory effect is more conspicuous at a lower temperature,
because the reservoir correlation functions decay more slowly with time.
Moreover, $I_L(t) = I_R(t)$ holds for all times because the two reservoirs are actually
equivalent in the absence of bias voltage and with the symmetric couplings ($\Gamma_L = \Gamma_R$).

Despite some latest progress, 
the HEOM approaches become exceedingly memory-consuming for ultra-low temperatures.
This is because, with the conventional (Matsubara or Pad\'{e}) spectral decomposition schemes, 
a large number of memory basis functions are required to
accurately unravel the reservoir correlation functions.
Therefore, in \Fig{fig2} we only present the HEOM results
for the high temperature $T_L = T_R = 1.0$.
In Paper~I, we have proved the MAS-SEOM is formally equivalent to the sim-HEOM approach,
and the interference auxiliary density operators (ADOs) omitted in the sim-HEOM
would affect the description of strongly correlated states such as the Kondo states.\cite{han2018exact}
Nevertheless, it is observed that the results of MAS-SEOM shown in \Fig{fig2}(a) and (b)
agree remarkably with those of the full HEOM.
This is because at the high temperature the formation of strongly correlated states
in the interacting AIM is suppressed by the thermal fluctuations,
and thus the simplified and full HEOM approaches yield almost the same results.

\textit{Voltage driven dynamics}:
Consider a single-level AIM with the electron-hole symmetry, i.e., $\epsilon_\uparrow =\epsilon_\downarrow = -U/2$.
The impurity level is initially half-filled with $(\rho_0)_{ij}=\frac{1}{4}\delta_{ij} \, \{i,j=1, \ldots,4 \}$.
At the time $t=0^+$ the impurity-reservoir couplings are turned on.
If the relaxation dynamics proceeds in the absence of bias voltage,
the impurity level will stay half-filled, and there is no apparent
electron transfer going on between the impurity and the reservoirs.
Consequently, we have $n_s(t) = n_s(0) = 0.5$ and $I_L(t) = I_R(t) = 0$ (data not shown).
Instead, when a constant bias voltage is applied asymmetrically
across the two reservoirs at $t > 0$, i.e., $V_L = - V_R$,
the impurity level remains half-filled, and the electric currents
in response to the bias voltage are also asymmetric, $I_L(t) = -I_R(t)$;
see \Fig{fig3}, because of the conservation of electrons in the total AIM.

For the relatively low temperature $T_L = T_R = 0.01$ studied in \Fig{fig3},
the results of MAS-SEOM still agree closely with those of the full HEOM.
This is somewhat surprising because Kondo states are expected to form
in the studied interacting AIM, and thus the results of MAS-SEOM (or sim-HEOM)
are expected to deviate from the full HEOM.
Indeed, such derivation would be observed if the MAS-SEOM was let to
propagate into the long-time regime, i.e., after the Kondo states are completely established.
However, it is difficult to reach the long-time regime
with the current implementation of MAS-SEOM,
because usually the number of trajectories ($N_{\rm traj}$)
required to achieve fully converged results is usually too large;
see \Sec{subsec:converge} for details.

We now elaborate on the numerical efficiency of the MAS-SEOM approach.
For an AIM consisting of $N_o$ impurity levels with $N_s$ spin directions
and $N_\alpha$ reservoirs, the number of pseudo-Fock-states in the
auxiliary space $S$ is $3^{2N_o N_s}$,
and the dimension of the impurity's Hilbert space is $(2N_s)^{N_o}$.
Therefore, for the single-level AIM examined in this section,
we have $N_o = 1$, $N_s=2$, $N_\alpha=2$, and thus the stochastic
reduced density matrix $\trhos$ is represented by a set of $81$ matrices
of the size $4\times 4$.
In contrast, with the HEOM method,
the width and depth of the hierarchy cannot be smaller than $M = 13$ and $L=4$,
to ensure an accurate unraveling of the reservoir memory
at the relatively low temperature $T_L = T_R = 0.01$.
This leads to a hierarchy of 2782131 ADOs of the size $4\times 4$.
Therefore, the cost of computer memory requested by the MAS-SEOM approach
is trivial as compared to that by the HEOM.
Apart from this, the trajectory-based algorithms make it easy to
do parallel computations with the MAS-SEOM approach.

\subsection{Convergence of MAS-SEOM} \label{subsec:converge}

%
\begin{figure}[t]
\centering
\includegraphics[width=\columnwidth]{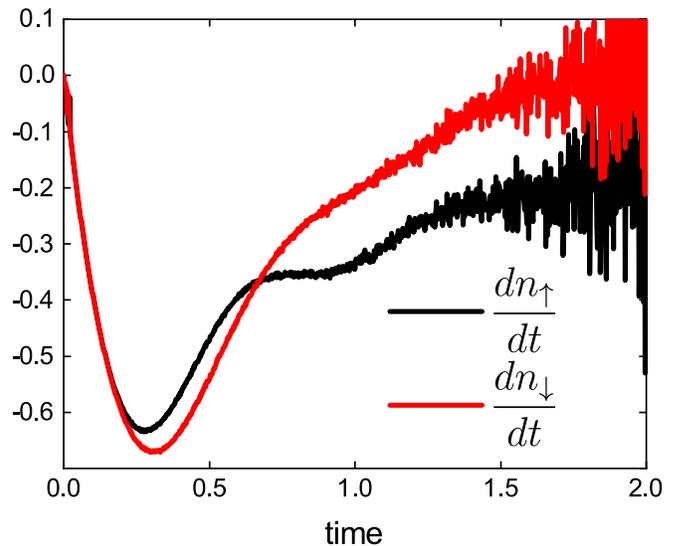}
\caption{Evolution of $dn_s/dt$ during the relaxation dynamics
of a single-level AIM at the temperature $T_L=T_R=0.1$.
Here, $n_s(t)$ is calculated by \Eq{operator}, and its time derivative
is computed by using the finite difference method.
The impurity level is initially doubly occupied, i.e., all the elements
of $\rho_0$ are zero except $(\rho_0)_{44}=1$.
The parameters of the AIM are: $\epsilon_{\uparrow}=0.5$, $\epsilon_{\downarrow}=-0.5$,
$U=5.0$, $\Gamma_L=\Gamma_R=0.5$, $\Omega=0$, and $W=5.0$;
see the main text in \Sec{subsec:accuracy} for the description of units.
The Euler-Maruyama algorithm is employed for the propagation of the MAS-SEOM,
with $dt=0.002$, $\lambda=1.0$, and $N_{\rm traj}=5 \times 10^7$.
}\label{fig4}
\end{figure}

\begin{figure}[t]
\centering
\includegraphics[width=\columnwidth]{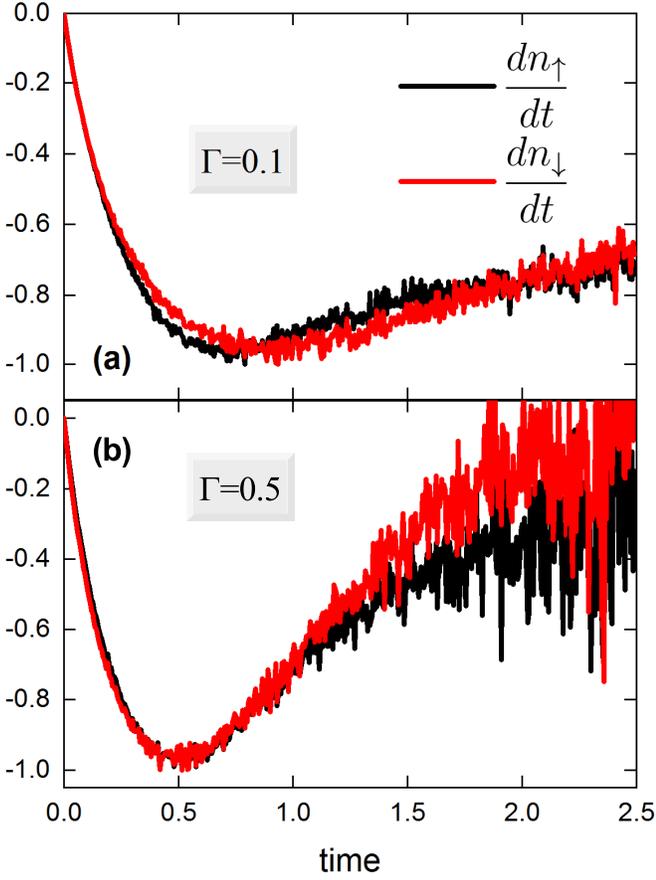}
\caption{Evolution of $dn_s/dt$ during the relaxation dynamics of
a single-level impurity coupled to a single electron reservoir,
with the coupling strength being (a) $\Gamma = 0.1$ and (b) $\Gamma = 0.5$.
The impurity level is initially doubly occupied, i.e., all the elements
of $\rho_0$ are zero except $(\rho_0)_{44}=1$.
The parameters of the AIM are: $\epsilon_{\uparrow}=0.5$, $\epsilon_{\downarrow}=-0.5$,
$U=2.0$, $\Omega=0$, $W=5.0$, and $T=0.5$;
see the main text in \Sec{subsec:accuracy} for the description of units.
The Euler-Maruyama algorithm is employed for the propagation of the MAS-SEOM,
with $dt=0.005$, $\lambda=1.0$, and $N_{\rm traj}=5 \times 10^5$. }
\label{fig5}
\end{figure}

\begin{figure}[t]
\centering
\includegraphics[width=\columnwidth]{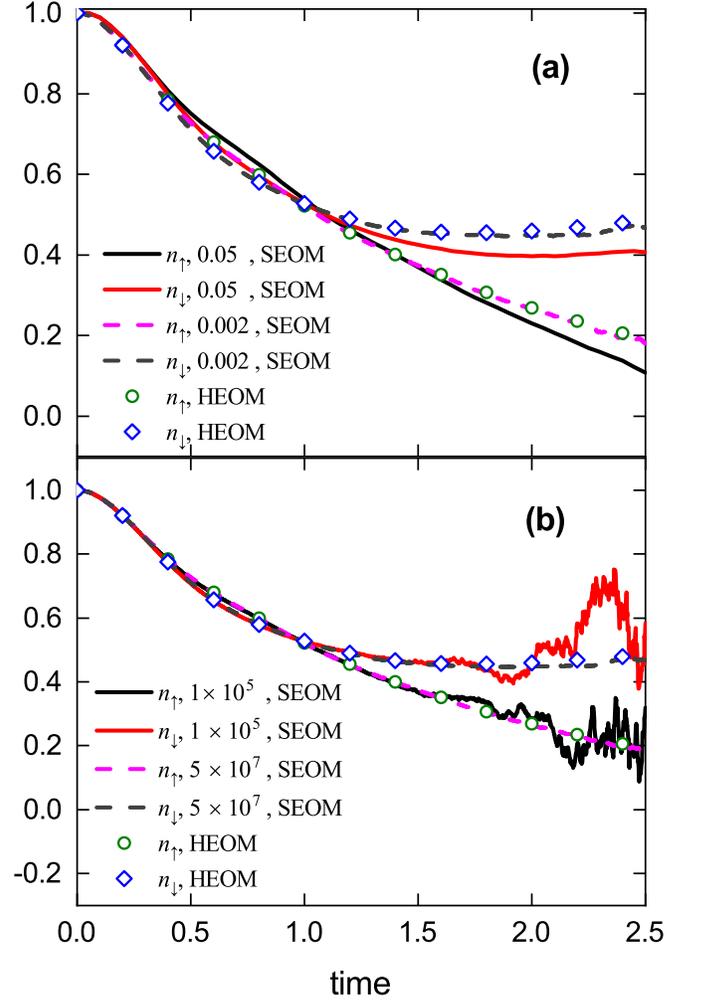}
\caption{Evolution of $n_s(t)$ during the relaxation dynamics of
a single-level AIM calculated with (a) different $dt$ and the same $N_{\rm traj}=5\times 10^7$
and (b) different $N_{\rm traj}$ and the same $dt=0.002$.
The impurity level is initially doubly occupied, i.e., all the elements
of $\rho_0$ are zero except $(\rho_0)_{44}=1$.
The parameters of the AIM are: $\epsilon_{\uparrow}=0.5$, $\epsilon_{\downarrow}=-0.5$,
$U=5.0$, $\Gamma_L=\Gamma_R=0.5$, $T_L=T_R=0.1$, $\Omega=0$, and $W=5.0$;
see the main text in \Sec{subsec:accuracy} for the description of units.
The Euler-Maruyama algorithm is employed for the propagation of the MAS-SEOM with $\lambda=1.0$.
The results of the full HEOM are also displayed as a reference for comparison.}
\label{fig6}
\end{figure}

In practical calculations, the MAS-SEOM is solved by generating a number of
quantum trajectories, with each trajectory serving as
a specific sample of the involving stochastic c-number noises.
Therefore, it is crucial that the group of trajectories accessed explicitly
in the calculation form an abundant sampling of all the random fields,
so that the statistical average of $\trhos$ and other key properties
can be obtained accurately.

As discussed in \Sec{sec:stability}, all the stochastic noises involved in
the MAS-SEOM of \Eq{Eq1} are bounded in the asymptotic limit.
This means that, in principle the correct statistical average can be achieved
as long as the number of trajectories ($N_{\rm traj}$) is sufficiently large.
However, from the analytic form of MAS-SEOM, it is clear that
the number of stochastic fields $\{v_{j\alpha st}, w_{j\alpha st}\}$
keeps increasing as the dissipative dynamics proceeds.
Therefore, it is expected that a much larger $N_{\rm traj}$ is needed
to yield fully converged results in the long-time regime than in the transient regime.
This is to be elucidated in this subsection.
Moreover, the numerical convergence of the MAS-SEOM may depend on other
aspects, such as the strength of impurity-reservoir couplings,
the length of time steps, the reservoir temperature, etc.
We will also examine the influence of these aspects.

To assess the numerical convergence of the MAS-SEOM, we explore
how the stochastic variance of the calculated $n_s(t)$ varies with time.
Instead of visualizing the values associated with individual trajectories
which have a rather scattered distribution,
we examine $dn_s/dt$ versus $t$.
While the averaged $n_s(t)$ is evaluated via \Eq{operator},
the time derivative is computed by using the finite difference method.
If the resulting $n_s(t)$ is fully converged, $dn_s/dt$ versus $t$ will be a smooth line;
otherwise the line will exhibit large oscillations, indicating
that the averaged $n_s(t)$ is subject to a large stochastic uncertainty.

Figure~\ref{fig4} depicts the variations of $dn_\uparrow/dt$ and $dn_\downarrow/dt$
during the relaxation dynamics of a single-level AIM.
Apparently, while the both lines are quite smooth in the short-time regime,
they start to oscillate after some time and the amplitudes of oscillations
increase gradually with time.
This indicates that the stochastic variance keeps growing
as the stochastic simulation proceeds.
As mentioned above, with the number of stochastic fields increasing with time,
the preset trajectories will gradually become inadequate for sampling
all the stochastic fields.

In the following, we discuss the influence of various aspects
on the convergence of the results of MAS-SEOM.

\textit{Influence of $\Gamma_\alpha$}:
From \Eqs{eq_omega} and \eqref{ct-zero-temp-1}, the amplitude of
the random field $w_{j\alpha st}$ is proportional to the strength
of impurity-reservoir coupling $\Gamma_\alpha$.
A stronger coupling $\Gamma_\alpha$ will thus lead to
a larger stochastic variance of $\trhos$.
Figure~\ref{fig5} depicts the evolution of $dn_s/dt$
during the relaxation dynamics of a single-level impurity
coupled to a single reservoir with different coupling strength $\Gamma$.
Apparently, at a same time, the larger $\Gamma$
indeed gives rise to more conspicuous oscillations in $dn_s/dt$.

\textit{Influence of $dt$ and $N_{\rm traj}$}:
To obtained converged results, it is important that the
time step is small enough so that the stochastic integrator is convergent.
For instance, \Fig{fig5}(a) compares $n_s(t)$ calculated by
employing the weak first-order Euler-Maruyama algorithm
with different time increment $dt$.   
Clearly, a too large $dt$ will yield $n_s(t)$ that
deviates significantly from the correct values.
Moreover, \Fig{fig5}(b) demonstrates that a larger $N_{\rm traj}$
is needed to acquire converged results at a longer time.

%
\begin{figure}[t]
\centering
\includegraphics[width=\columnwidth]{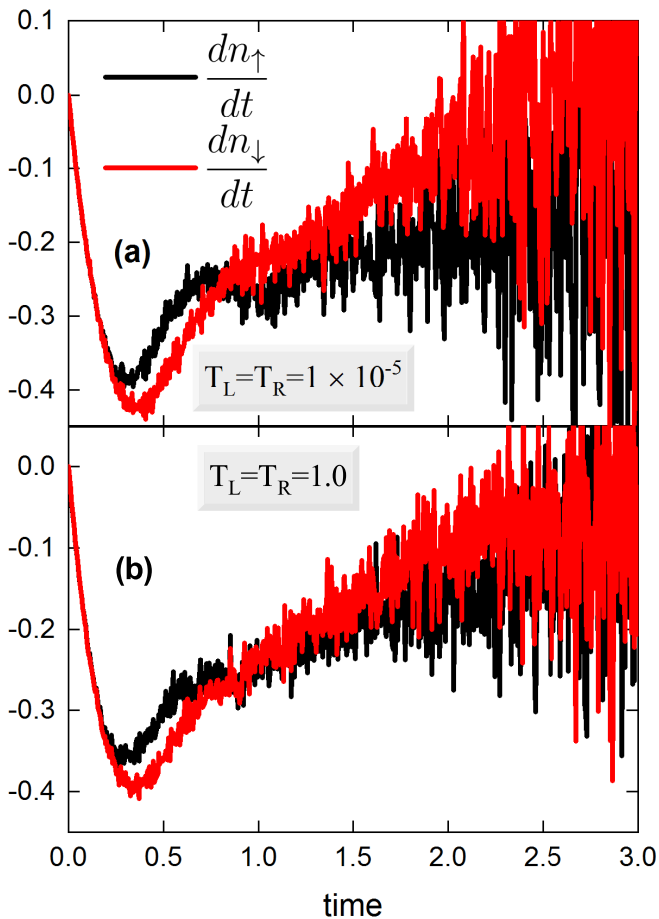}
\caption{Evolution of $dn_s/dt$ during the relaxation dynamics
of a single-level AIM at the temperature of
(a) $T_L=T_R=1.0\times 10^{-5}$ and (b) $T_L=T_R=1.0$.
The initial condition and the parameters of AIM are identical to those adopted for \Fig{fig2}.
The Euler-Maruyama algorithm is employed for the propagation of the MAS-SEOM
with $dt=0.005$, $\lambda=1.0$, and $N_{\rm traj}=1 \times 10^5$.}
\label{fig7}
\end{figure}

\textit{Influence of $T$}:
From the analysis in \Sec{sec:stability},
a more slowly decaying reservoir correlation function corresponds to
a larger amplitude of the random field $w_{j\alpha st}$.
Thus, a lower temperature will give rise to a larger stochastic variance
in the resulting $\trhos$.
Figure~\ref{fig7} compares the evolution of $dn_s/dt$ at very different temperatures.
In the long-time regime, the oscillations in $dn_s/dt$
at the ultra-low temperature are somewhat more pronounced
than at the high temperature, but the difference is not drastic.
Therefore, we see that the MAS-SEOM is indeed a favorable method
for the study of low-temperature quantum dissipative dynamics.

\textit{Influence of $\lambda$}:
In the MAS-SEOM of \Eq{Eq1}, the parameter $\lambda$ tunes
the relative amplitudes of the instantaneous random fields $\{v_{j\alpha st}\}$
and the memory-convoluted fields $\{w_{j\alpha st}\}$.
Thus, in principle there exists a $\lambda$ that results in
most balanced random fields, and thus leads to an
optimal convergence for the stochastic simulation.
However, in practice we have not observed any substantial
improvement in the convergence by varying the value of $\lambda$.
A more careful numerical analysis is needed to clarify this issue.

\textit{Influence of stochastic integrator}:
In this work, the weak first-order Euler-Maruyama algorithm is
employed to propagate the MAS-SEOM.
Higher-order stochastic integrators have been proposed,\cite{Klo92,jacobs2010stochastic,yan2016stochastic,ullah2017monte,Sun19136766}
which are expected to yield much improved convergence.
The sophisticated stochastic integrators will also enhance
the efficiency of the stochastic simulation, because
they allow for the use of much larger time steps.
We leave the implementation of the higher-order algorithms
for future work.

\section{CONCLUDING REMARKS AND PERSPECTIVES}  \label{sec:conclusion}

In this paper, we present the numerical implementation of the MAS-SEOM method
for a single-level impurity coupled to two electron reservoirs.
The direct stochastic simulations for both the relaxation and voltage-driven
dynamics of the AIM are demonstrated, with detailed discussions
on the accuracy, efficiency and convergence of the MAS-SEOM.
The presented results clearly advocate the MAS-SEOM as a promising
candidate for the study of non-equilibrium dynamics of QIS.

From the given numerical examples, in the short-time regime the MAS-SEOM
yields accurate results that agree remarkably with those of the full HEOM;
whereas in the long-time regime, the stochastic variance of $\trhos$ grows rapidly,
and it requires to use too many trajectories to attain fully converge results
with the low-order Euler-Maruyama algorithm.
Therefore, the development and application of higher-order stochastic integrators
which allow for the use of larger time steps
are essentially important to make the MAS-SEOM practical.
Nevertheless, the MAS-SEOM has shown great potential in the study of
fermionic dissipative dynamics at ultra-low temperatures, which is beyond
the capability of the present HEOM method.

In Paper~I, it has been proved that the MAS-SEOM is equivalent to
the sim-HEOM formalism in which the interference ADOs that are
important for the description of strongly correlated states are left out.
For the relaxation dynamics starting from a decoupled initial state,
the discrepancies between the results of MAS-SEOM (or sim-HEOM) and
those of the full HEOM are expected to arise in the long-time regime.
To reduce such discrepancies, a more sophisticated mapping strategy
for representing the time-dependent AGFs is called for.

To summarize, the MAS-SEOM method lays a foundation for the
direct stochastic simulation of fermionic dissipative dynamics.
Admittedly, there is still much room for improvement in
many aspects of the MAS-SEOM approach, including its accuracy,
efficiency, convergence, and applicability.
Many existing algorithms and techniques adopted in the stochastic
simulation of bosonic open systems can be transferred
straightforwardly to the study of fermionic QIS,
such as the construction of color noises with preset cross-correlations,
the use of high-order stochastic integrators,
and the parallel computing techniques.
With the future improvements, the SEOM method has great potentials
to become a useful theoretical tool for the investigation of
strongly correlated QIS.

\acknowledgments
Support from the Ministry of Science and Technology of China
(Grants No.\ 2016YFA0400900 and No.\ 2016YFA0200600),
the National Natural Science Foundation of China
(Grants No.\ 21973086, No.\ 21573202, No.\ 21633006, No.\ 21973036 and
No.\ 21903078), and the Ministry of Education of China (111 Project Grant No.\ B18051)
is gratefully acknowledged.
V.Y.C. was supported by the U.S. Department of Energy, Office of Science, Basic Energy Sciences,
Materials Sciences and Engineering Division, Condensed Matter Theory Program.

\appendix*

\section{Application of MAS-SEOM to a closed two-level system} \label{append:two_level}
%

\begin{figure*}[t]
\centering
\includegraphics[width=2\columnwidth]{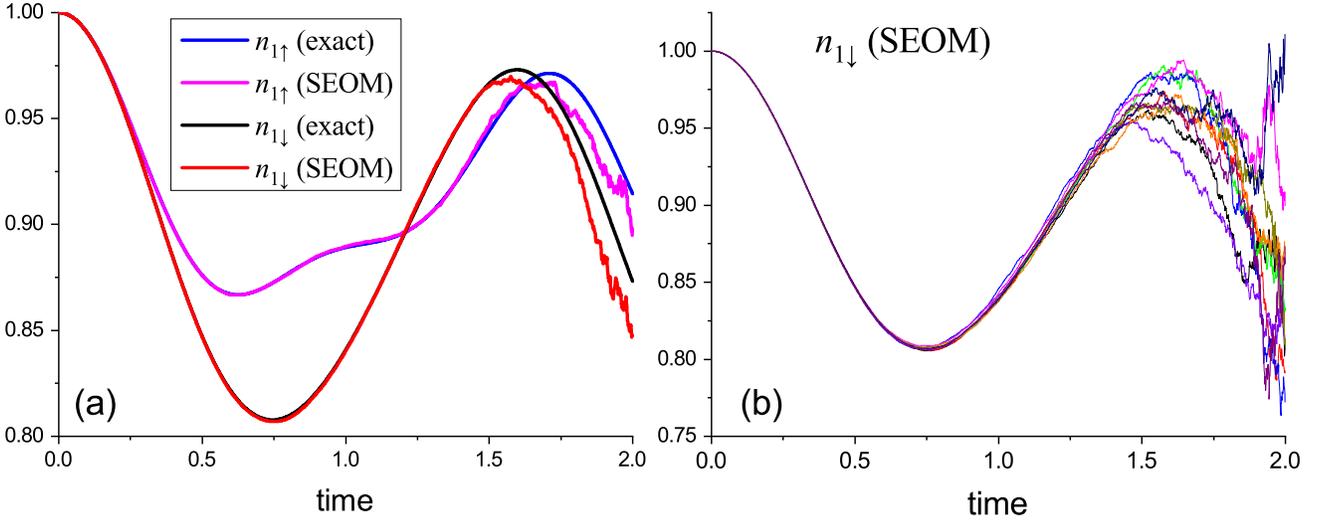}
 \caption{(a) Evolution of $n_{1s}(t)$ during the relaxation dynamics of a closed two-level system.
  (b) Each line is the averaged $n_{1\downarrow}(t)$ over $5\times 10^5$ trajectories calculated by the MAS-SEOM method.
  The level-1 is initially doubly occupied (all the elements of $\rho_1^0$ are zero except $(\rho_1^0)_{44}=1$),
  while the level-2 is initially vacant (the only nonzero element of $\rho_2^0$ is $(\rho_2^0)_{11}=1$).
  The parameters of the two-level system are: $\epsilon_{1\uparrow} = 0.5$, $\epsilon_{1\downarrow} = -0.5$,
  $U_1=5.0$, $\epsilon_{2\uparrow} = \epsilon_{2\downarrow} = U_2 = 0$, and $\Gamma = 1.0$;
  see the main text in \Sec{subsec:accuracy} for the description of units.
  The MAS-SEOM of \Eq{Eq1-rho1} is propagated by using the Euler-Maruyama algorithm with
  $dt = 0.001$, $\lambda=1.0$, and $N_{\rm traj} = 5\times 10^6$.
  The exact results obtained by solving the \Sch equation for the total two-level system
  are displayed in (a) as a reference for comparison.}
  \label{fig8}
\end{figure*}
%

For illustrative purpose, in the following we show that the MAS-SEOM can be applied to
a simple toy model, a closed two-level system, to describe the reduced dynamics of
one of the two levels.

The closed two-level system is described by the Hamiltonian
\begin{equation}
 \Ht = H_1 + H_2 + H_{\rm int} \ , \label{hamil-2L}
\end{equation}
where $H_\nu = \epsilon_{\nu\uparrow} \hat{n}_{\nu\uparrow} + \epsilon_{\nu\downarrow} \hat{n}_{\nu\downarrow}
+ U_\nu \hat{n}_{\nu\uparrow}\hat{n}_{\nu\downarrow}$  ($\nu=1,2$),
and $H_{\rm int}= \Gamma \sum_s \hat{c}^\dag_{1s}\hat{c}_{2s} + \hat{c}^\dag_{2s}\hat{c}_{1s}$
is coupling Hamiltonian between the two levels.
The quantum dynamics of the two level system is exactly described
by the \Sch equation for the total density matrix $\rhot$:
\be \label{eom-rhot}
  \dot{\rho}_{_{\rm T}} = -i [\Ht, \rhot].
\ee

Alternatively, as described in Paper~I, the dynamics of the two levels
can be formally decoupled as $\rhot = \la \rho_1 \rho_2 \ra$
by introducing the time-dependent AGFs $\{\eta_{j st}, \bar{\eta}_{j st}\}$
($j=1,\ldots,4$), where $\rho_\nu$ is the stochastic reduced
density matrix of the $\nu$th level. With the initial condition
$\rhot(t_0) = \rho_1^0 \rho_2^0$, the formally exact SEOM for $\rho_1$ and $\rho_2$
can be derived as
\begin{align}
\dot{\rho}_1 = & -i[H_1, \rho_1] + \lambda^{\frac{1}{2}} \!
\sum_{s=\uparrow,\downarrow} \big[ e^{-\frac{i\pi}{4}} \big( \hat{c}_{1s}^\dag\,\eta_{1st} + \bar{\eta}_{2st}\, \hat{c}_{1s} \big) \rho_1 \nl
&\qquad +  e^{\frac{i\pi}{4}} \rho_1  \big(\hat{c}_{1s}^\dag\,\eta_{3st} + \bar{\eta}_{4st}\,\hat{c}_{1s} \big) \big],  \label{rho_1} \\
\dot{\rho}_2 = & -i[H_2, \rho_2] + \lambda^{-\frac{1}{2}}\Gamma \sum_{s=\uparrow,\downarrow}
 \big[ e^{-\frac{i\pi}{4}} \big( \bar{\eta}_{1st}\, \hat{c}_{2s}  +  \hat{c}_{2s}^\dag \, \eta_{2st}  \big) \rho_2 \nl
&\qquad +  e^{\frac{i\pi}{4}} \rho_2  \big( \bar{\eta}_{3st}\,\hat{c}_{2s}^\dag  + \hat{c}_{2s}^\dag\, \eta_{4st} \big) \big] \label{rho_2}.
\end{align}
Define $\tilde{\rho}_1 \equiv \rho_1 {\rm tr}_2(\rho_2)$,
so that the quantum trajectories of $\tilde{\rho}_1$ are equally weighted,
and the averaged reduced density matrix of level-1 is $\bar{\rho}_1 = \la \tilde{\rho}_1\ra$.

If the level-2 is non-interacting, i.e., $U_2 = 0$,
${\rm tr}_2(\rho_2)$ can be evaluated by using the Magnus expansion\cite{tannor2007introduction}
and the Baker-Campbell-Hausdorff formula,\cite{Gre96} as follows,
\begin{align}
  \text{tr}_{2} ( \rho_2 )  &= \exp \Big\{  \sum_{s = \uparrow,\downarrow}
   \int_{t_0}^t  d\tau \, \Big[ \big(\bar{\eta}_{1s \tau} - i\bar{\eta}_{3s \tau} \big) g^{-}_{s \tau} \nl
   & \qquad \qquad \qquad + \big(\eta_{2s \tau} - i\eta_{4s \tau} \big)  g^{+}_{s \tau} \Big] \Big \} ,   \label{trace-rho_2}
\end{align}
with
\begin{align}
 g^{-}_{s t} &= \lambda^{-1} \Gamma^2  \int_{t_0}^t  d\tau \, e^{-i\epsilon_{2s}(t-\tau)}
     \left( n_{2s}^0\, \eta_{4s\tau} - i \bar{n}_{2s}^0 \, \eta_{2s\tau} \right),  \nl
 g^{+}_{s t}  &= \lambda^{-1} \Gamma^2 \int_{t_0}^t  d\tau\,  e^{\,i\epsilon_{2s}(t-\tau)}
    \left( \bar{n}_{2s}^0\, \bar{\eta}_{3s\tau} - i n_{2s}^0 \, \bar{\eta}_{1s\tau} \right). \label{gb-2L-1}
\end{align}
Here, $n_{2s}^0 = {\rm tr}_2 \big(\hat{c}^\dag_{2s}\hat{c}_{2s}\,\rho_2^0\big)$
and $\bar{n}_{2s}^0 = {\rm tr}_2 \big(\hat{c}_{2s}\hat{c}^\dag_{2s}\,\rho_2^0\big) = 1-n_{2s}^0$
are the electron and hole occupation numbers on the level-2 at $t_0$, respectively.
After a Grassmann Girsanov transformation,\cite{han2019fermionic}
we achieve the SEOM for $\tilde{\rho}_1$ as follows:
\begin{align}  \label{seom-rho1}
\dot{\tilde{\rho}}_1 &=  -i[H_1, \tilde{\rho}_1] + \lambda^{\frac{1}{2}} \sum_{s=\uparrow,\downarrow}
\Big[\, e^{-\frac{i\pi}{4}} \big\{ \hat{c}_{1s}^\dag\,g^{-}_{st} -g^{+}_{st}\,\hat{c}_{1s}, \, \tilde{\rho}_1 \big\} \nl
&\qquad \qquad  + e^{-\frac{i\pi}{4}} \big( \hat{c}_{1s}^\dag\, \eta_{1st} + \bar{\eta}_{2st} \,\hat{c}_{1s} \big) \rho_1 \nl
&\qquad \qquad  + e^{\frac{i\pi}{4}} \rho_1  \big(  \hat{c}_{1s}^\dag\, \eta_{3st} + \bar{\eta}_{4st}\, \hat{c}_{1s} \big) \Big] .
\end{align}
By using the MAS mapping of \Eq{map_eq1}, we arrive at the following MAS-SEOM
\begin{align} \label{Eq1-rho1}
\dot{\tilde{\rho}}_1 &= -i[H_1, \tilde{\rho}_1] + \lambda^{\frac{1}{2}}  \sum_{s=\uparrow,\downarrow}
 \Big[ e^{-\frac{i \pi}{4}} \big(\hat{c}^\dagger_{1s} \, Y_{1 s} + Y_{2 s}\, \hat{c}_{1s} \big) \tilde{\rho}_1  \nl
&\qquad \qquad
+ e^{\frac{i \pi}{4}} \tilde{\rho}_1 \big(\hat{c}^\dagger_{1s}\, Y_{3 s} + Y_{4 s}\, \hat{c}_{1s} \big) \Big],
\end{align}
where
\begin{align} \label{parameters-rho1}
 Y_{1 s} &\equiv v_{1 st}\, X^-_{1s} + \tilde{g}^-_{ st}\,,   \ \ \ \  Y_{2 s} \equiv v_{2 st}\, X^+_{2s} - \tilde{g}^+_{ st}\,,  \nl
 Y_{3 s} &\equiv v_{3 st}\, X^-_{1s} - i \tilde{g}^-_{ st}\,,  \ \ \ Y_{4 s} \equiv v_{4 st}\, X^+_{2s} + i \tilde{g}^+_{ st}\,,
\end{align}
and
\begin{align} \label{def-tg-rho1}
  \tilde{g}^-_{s t} &=\lambda^{-1} \Gamma^2 \int_{t_0}^t d\tau\,  e^{-i\epsilon_{2s}(t-\tau)}
  \left( n_{2s}^0 v_{4 s\tau} -i \bar{n}_{2s}^0\, v_{2 s\tau}\right) X^-_{2s}, \nl
  \tilde{g}^+_{s t} &=\lambda^{-1} \Gamma^2 \int_{t_0}^t d\tau\,  e^{i\epsilon_{2s}(t-\tau)}
  \left(\bar{n}_{2s}^0 v_{3 s\tau} -in_{2s}^0 v_{1 s\tau}\right) X^+_{1s}.
\end{align}

We emphasize that the MAS-SEOM of \Eq{Eq1-rho1} is formally exact,
as long as the level-2 is non-interacting.
This is because the ``reservoir'' is a single level (level-2)
and the reservoir correlation function is a single exponential function,
cf. \Eq{def-tg-1} and \Eq{def-tg-rho1}.
Therefore, the HEOM formalism which is formally equivalent to \Eq{seom-rho1}
does not involve any interference ADO,\cite{han2018exact}
and thus the sim-HEOM and the MAS-SEOM of \Eq{Eq1-rho1} are also formally exact.

Figure~\ref{fig8}(a) depicts the evolution of $n_{1s}(t) = {\rm tr}_1\big(\hat{n}_{1s} \la \tilde{\rho}_1 \ra\big)$
after switching on the inter-level coupling at $t = 0$.
The results of MAS-SEOM are compared against the exact solution obtained from \Eq{eom-rhot}.
Apparently, in the short-time regime, the predictions of MAS-SEOM
agree perfectly with the exact results.
However, at $t > 1.5$ the results of MAS-SEOM start to deviate from the exact lines.
As explained in \Sec{sec:stability}, such deviations arise because
the ``reservoir'' correlation function does not decay with time,
and thus the amplitudes of the memory-convoluted noises keep growing.
Consequently, the MAS-SEOM of \Eq{Eq1-rho1} is asymptotically unstable
for the two-level system.
It is clearly seen in \Fig{fig8}(b) that the stochastic variance
begins to diverge from $t > 1.5$.


\end{document}